\documentclass[12pt, draftclsnofoot, onecolumn]{IEEEtran}
\usepackage{epsfig}
\usepackage{amssymb}
\usepackage[font=footnotesize]{caption} % small
\usepackage{latexsym,graphicx}
\usepackage{longtable}
\usepackage{algpseudocode} 
\usepackage{algorithm}
\usepackage{array}
\usepackage{mathrsfs}
\usepackage{color}
\usepackage{amsmath,graphicx}
\usepackage{relsize}
\usepackage{epsfig}
\usepackage{setspace}
\usepackage{cite}
\usepackage{hyperref}
\usepackage{amssymb}
\usepackage{caption}
\usepackage{subcaption}
\usepackage[normalem]{ulem}
\usepackage{soul,xcolor}
\def\BibTeX{{\rm B\kern-.05em{\sc i\kern-.025em b}\kern-.08em
    T\kern-.1667em\lower.7ex\hbox{E}\kern-.125emX}}

 \makeatletter
 \let\NAT@parse\undefined
 \makeatother

\begin{document}
%\title{Square Deviation Based Symbol-Level Selective Relaying for Full-Duplex Cooperative Network} 
\title{Symbol-Level Selective Full-Duplex Relaying with Power and Location Optimization} 
\author{Jiancao Hou, \IEEEmembership{Member, IEEE,} Sandeep Narayanan, \IEEEmembership{Member, IEEE,} \\Na Yi, \IEEEmembership{Member, IEEE,} Yi Ma, \IEEEmembership{Senior Member, IEEE,} \\and Mohammad Shikh-Bahaei, \IEEEmembership{Senior Member, IEEE}% <-this % stops a space
%\thanks{This work was partly funded by the ICT-248894 WHERE 2.}
% <-this % stops a space
\thanks{This work was supported by the Engineering and Physical Science Research Council (EPSRC) through the SENSE grant EP/P003486/1.

J. Hou, S. Narayanan and M. Shikh-Bahaei are with the Department of Informatics, King's College London, London, United Kingdom, WC2R 2LS. E-mail: \{jiancao.hou, sandeep.kadanveedu, m.sbahaei\}@kcl.ac.uk.

N. Yi and Y. Ma are with the Institute for Communication Systems, University of Surrey, Guildford, United Kingdom, GU2 7XH. E-mail: \{n.yi, y.ma\}@surrey.ac.uk.}
}
\markboth{} {Shell \MakeLowercase{\textit{et al.}}: Bare Demo of
IEEEtran.cls for Journals}\maketitle

\begin{abstract}
%Summary of your paper: brief introduction, motivation, and contributions.
In this paper, a symbol-level selective transmission for full-duplex (FD) relaying networks is proposed to mitigate error propagation effects and improve system spectral efficiency. The idea is to allow the FD relay node to predict the correctly decoded symbols of each frame, based on the generalized square deviation method, and discard the erroneously decoded symbols, resulting in fewer errors being forwarded to the destination node. Using the capability for simultaneous transmission and reception at the FD relay node, our proposed strategy can improve the transmission efficiency without extra cost of signalling overhead. In addition, targeting on the derived expression for outage probability, we compare it with half-duplex (HD) relaying case, and provide the transmission power and relay location optimization strategy to further enhance system performances. The results show that our proposed scheme outperforms the classic relaying protocols, such as cyclic redundancy check based selective decode-and-forward (S-DF) relaying and threshold based S-DF relaying in terms of outage probability and bit-error-rate. Moreover, the performance with optimal power allocation is better than that with equal power allocation, especially when the FD relay node encounters strong self-interference and/or it is close to the destination node. 
\end{abstract}


\IEEEpeerreviewmaketitle
%%%%%%%%%%%%%%%%%%%%%%%%%%%%%%%%%%%%%%%%%%%%%%%%%%%%%%%%%%%%%%%%%%%%%%
 
                         %%%I. Introduction%%% 

%%%%%%%%%%%%%%%%%%%%%%%%%%%%%%%%%%%%%%%%%%%%%%%%%%%%%%%%%%%%%%%%%%%%%%
\section{Introduction}
%Para. 1 introduces the background related to your research.
Cooperative relaying has attracted much attention in recent years due to its capability to serve as a virtual multi-antenna system to combat fading and improve spectral efficiency \cite{Jamal2015}. In general, based on its method of operation, cooperative relaying can be classified into three categories \cite{Laneman2004,Lai2006}: 1) amplify-and-forward (AF) relaying, where the relay node simply amplifies the received signal and forwards it to the destination node; 2) decode-and-forward (DF) relaying, where the relay node decodes the received signal and forwards the regenerated signal to the destination node; 3) compress-and-forward (CF) relaying, where the relay node compresses and quantizes its received signal and then forwards it to the destination node. In these three protocols, DF relaying simplifies the power control and allows for reprocessing of the decoded signal at the relay node \cite{Al-Habian2011}. However, such protocol may encounter error propagation effects, which can degrade the system performance \cite{Kwon2010a}.

%Para. 2 explains the motivations for your research.
On the other hand, in order to fully exploit spatial diversity gain and avoid co-channel interference, conventional relaying protocols normally work in half-duplex (HD) mode, where the HD relay node either receives or transmits data symbols at any given time-instant. However, HD relaying suffers from multiplexing loss since the transmission of one data frame occupies two successive time slots. To recover this loss, many relaying protocols have been proposed in the literature, and one of their notable examples is named two-path successive relaying \cite{Oechtering2004,Lehmann2016}. This scheme mimics full-duplex (FD) relaying and allows the source node to continuously transmit information data for every channel use, while two HD relay nodes alternately serve as transmitter and receiver to relay the source node's messages. In this case, diversity-multiplexing tradeoff (DMT) was usually used to identify the system performance, and the optimum DMT performance is achieved when both HD relay nodes can perfectly decode the messages sent from the source node\cite{Fan2007,Wicaksana2011}. Otherwise, the error propagation will degrade the system performance. 

Consider a system with only three nodes. Recent research works in \cite{Almradi2016,Sabharwal2014,Naslcheraghi2017} show that FD relaying has become feasible for simultaneous transmission and reception at the same frequency if the self-interference cancellation at the relay node can be well exploited \cite{Liu2015,Sim2017}. However, in practical environments, the self-interference effect cannot be cancelled perfectly for a variety of reasons, such as imperfect channel estimation and/or limited dynamic range of the analog-to-digital converter (A/D), which leads to having residual self-interference at the FD relay node\cite{Kim2015,Yu2015}. The presence of residual self-interference affects the decoding process at the FD relay node and may lead to error propagation effects. In order to mitigate these effects, the FD relay node can implement frame-level selective decode-and-forward (S-DF) strategies, such as cyclic-redundancy code (CRC) check based S-DF or signal-to-interference plus noise ratio (SINR) threshold based S-DF, where the relay node prevents it from forwarding if CRC fails or its SINR is below the predetermined threshold \cite{Olivo2016,Khafagy2015,Woradit2009}. However, both strategies result in diversity degradation since a single or a few error bits in a coded frame would hinder a significant number of correctly decoded bits to be forwarded to the destination node. Moreover, retransmission requests which aim to guarantee the perfect decoding may also cause spectral efficiency loss.

To compensate the loss of diversity gain and spectral efficiency, symbol-level selective methods have been proposed for HD relaying in \cite{Al-Habian2011,Kwon2010a,Yi2015}. In \cite{Al-Habian2011}, a log-likelihood ratio (LLR) based selection method was proposed, where the HD relay node first calculates LLR values of its decoded bits, and then compares them with a predetermined threshold to decide which bits should be forwarded to the destination node. The authors in \cite{Kwon2010a} proposed a similar LLR-based symbol-level selective transmission for the demodulation-and-forward relaying. Unlike the work in \cite{Al-Habian2011}, the received signal at the HD relay node in \cite{Kwon2010a} is only demodulated/re-modulated in a symbol-by-symbol manner without any decoding/re-encoding operations. It is worth noting that, to precisely select the qualified bits or symbols at the HD relay node, both approaches need to obtain appropriate LLR threshold. However, such threshold is hard to find in practice, especially for generalized modulated constellation schemes. The authors in \cite{Yi2015} then proposed an absolute difference based selection criteria that no preset threshold is needed. Specifically, the relay node first formulates the absolute difference between the re-constructed signal and the received signal, and then compares the absolute difference with the receive signal to identify whether the signal is detected correctly. To provide an accurate prediction, the modulation scheme of the selection method has to be limited to binary phase shift keying (BPSK). Moreover, for improving the spatial diversity gain at the destination node, all above three techniques require the relay node to inform the destination node about the positions of its discarded symbols of each frame at the cost of signalling overhead \cite{Kobravi2007}.

Motivated by the above discussion, a simple square deviation based symbol-level selective method for FD relaying is proposed in this paper to improve spectral efficiency and reliability. Our contributions can be summarized in the following points.

\begin{itemize}
\item First, a square deviation based symbol-level selection method is proposed for FD relaying networks, which aims to predict the correctly decoded symbols based on the square distance between the reconstructed symbols and the received signals after the linear detection process. Unlike the work in \cite{Yi2015}, by stating the problem as an integer square deviation problem, our proposed method is suitable for generalized modulation schemes \cite{Xu2013}. Moreover, with the selected linear detection processing on the received signal at the FD relay node, our proposed method can suppress channel fading, self-interference and noise effects simultaneously. In addition, different from the symbol-level selective methods in \cite{Al-Habian2011,Kwon2010a}, our proposed method avoids the complicated predetermined threshold setup, and saves additional signalling to the destination node about the spatial diversity combining.

\item Due to FD relaying, the destination node may encounter inter-frame interference. Unlike the work in \cite{Yu2015,Kwon2010} where the S-D link signal is treated as noise at the destination node, our proposed scheme utilizes the S-D link information during the detection/decoding process. In this case, a modified maximum \textit{a posteriori} (MAP) detector is proposed to mitigate inter-frame interference and improve the system spatial diversity gain.

\item As shown in \cite{Hou2015a,Nehra2010,Zarringhalam2009}, efficient resource allocation can help to improve system performances. In this paper, an outage probability based transmission power and relay location optimization is analysed. To achieve this, the outage probability expression for our proposed scheme is first derived, and then the optimal power allocation and relay location placement are demonstrated. Apart from that, we compare FD version with HD version of our proposed scheme. Following the theoretical analysis, computer simulations are provided to illustrate the advantage of our proposed scheme by comparing with classic relaying protocols. The results also show that the proposed power allocation outperforms the equal power allocation especially when the self-interference level is high. In addition, if the FD relay node has strong decoding capability, locating the FD relay node closer to the destination node leads to a better system performance. %in order to verify the results. 
\end{itemize}

The rest of the paper is organized as follows. Section II presents the system model of three nodes FD relaying network. Section III provides the proposed symbol-level selective transmission method at the FD relay node and decoding method at the destination node. The outage probability based power and location optimization is analysed in Section IV. Section V gives numerical and simulation results, and Section VI concludes the paper. Throughout this paper, the basic notations have been summarized in Tab.~\ref{tab1}. 
\begin{table}[t]
\center
{\small
\caption{Summarizes the basic notations in the paper}\label{tab1}
\begin{tabular}{ |p{2.4cm}||p{12.5cm}|}
%\hline
%\multicolumn{2}{|c|}{Notation List} \\
\hline
\textbf{Symbol} & \textbf{Usage}\\
\hline
$\mathcal{R}^{N}$, $\mathcal{C}^{N}$ & The set of real and complex $N$-tuples, respectively\\
\hline
$L$ & The total number of transmission frames \\
\hline
$M$ & The number of symbols per frame \\
\hline
$k$ & Channel coding rate \\
\hline
$\mathcal{R}(x)$, $\mathcal{I}(x)$  & The real and imaginary parts of a complex number $x$, respectively \\
\hline
$P_{\mathrm{S}},P_{\mathrm{R}}$ & Transmission powers with respect to the source and the FD relay nodes, respectively\\
\hline
$h_{i,j}(l)$ & Instantaneous channel coefficient between node $i$ and node $j$ in the $l^{\mathrm{th}}$ time slot\\
\hline
${x}^{(m)}_{\mathrm{S}}(l),x^{(m)}_{\mathrm{R}}(l)$ & The complex-valued $m^{\mathrm{th}}$ symbol of the $l^{\mathrm{th}}$ transmission frames from the source and the FD relay nodes, respectively\\
\hline
$d_{i,j}$ & Distance between node $i$ and node $j$ \\
\hline
$v$ & Path loss exponent\\
\hline
${v}^{(m)}_{R}(l)$ & The noise at the relay node for the $m^{\mathrm{th}}$ symbol of the $l^{\mathrm{th}}$ transmission frame \\
\hline
$\tilde{\mathbf{x}}^{(m)}_{\mathrm{S}}(l)$, $\tilde{\mathbf{x}}^{(m)}_{\mathrm{R}}(l)$  & The real-valued $m^{\mathrm{th}}$ symbol of the $l^{\mathrm{th}}$ transmission frame from the source and the FD relay nodes, respectively\\
\hline
$\varepsilon$ & Square deviation error\\
\hline
$c^{(m)}_{\mathrm{S},i}(l)$, $c^{(m)}_{\mathrm{R},i}(l)$ & The $i^{\mathrm{th}}$ bit that used to modulate the $m^{\mathrm{th}}$ symbol of the $l^{\mathrm{th}}$ frame from the source and the FD relay nodes, respectively\\
\hline
$\hat{\mathbf{x}}^{(m)}_{\mathrm{R}}(l)$ & The regenerated real-valued $m^{\mathrm{th}}$ symbol of the $l^{\mathrm{th}}$ frame at the FD relay node just before the proposed symbol-level selection operation \\
\hline
$\sigma^{2}_{h_{i,j}}$ & The variance of channel link between node $i$ and node $j$ \\
\hline
$\mathcal{P}_{\mathrm{S}}$ & The probability of a symbol is selected to be forwarded to the destination node\\
\hline
$\mathcal{P}_{\mathrm{C}}$ & The average probability of correctly predicted/forwarded symbols per frame  \\
\hline
$\mathcal{P}_{1}$ & The probabilities of $\mathcal{P}_{\mathrm{S}}$ on the condition that self-interference effect does exist  \\
\hline
$\mathcal{P}_{0}$ & The probabilities of $\mathcal{P}_{\mathrm{S}}$ on the condition that self-interference effect does not exist  \\
\hline
$R$ & Transmission rate at the source and the relay nodes \\
\hline
$G=([i,j])_{8}$ & Generator polynomial in octal form ($i$ and $j$ denote the values of the first and the second generators, respectively)\\
\hline
$\Gamma_{\mathrm{T}}$ & Predetermined SINR threshold for the threshold based S-DF scheme \\
\hline
$Z$ & The number of coded bits per symbol \\
\hline
\end{tabular}}
\end{table}

%Throughout the paper, lower case boldface symbols are used to denote column vectors (e.g. $\mathbf{a}$), while upper case boldface symbols are used to denote matrices (e.g. $\mathbf{A}$). $(\cdot)^{T}$ denotes the vector (or matrix) transpose. $\mathcal{C}^{N}$ and $\mathcal{R}^{N}$ denote the set of complex and real $N$-tuples, respectively. $\mathcal{R}(x)$ and $\mathcal{I}(x)$ denote the real value and imaginary value of complex number $x$, respectively. $\mathbf{I}_{M}$ denotes a $M\times M$ identity matrix. $\|\cdot\|$ denotes euclidean norm, $[\cdot]^{-1}$ denotes inverse of a matrix, and $\mathrm{diag}(\mathbf{a})$ denotes a diagonal matrix whose diagonal is vector $\mathbf{a}$.

%%%%%%%%%%%%%%%%%%%%%%%%%%%%%%%%%%%%%%%%%%%%%%%%%%%%%%%%%%%%%%%%%%%%%%
 
                         %%%II. System Model%%% 

%%%%%%%%%%%%%%%%%%%%%%%%%%%%%%%%%%%%%%%%%%%%%%%%%%%%%%%%%%%%%%%%%%%%%%
% Provide a comprehensive description of the system model which leads to your theoretical analysis. A block diagram is usually demanded to clarify your % presentation. Formulate the problem you want to resolve. Please note this system model should be much related to your simulations.
\section{System Model}
\begin{figure}[t] 
\begin{center}
\epsfig{figure=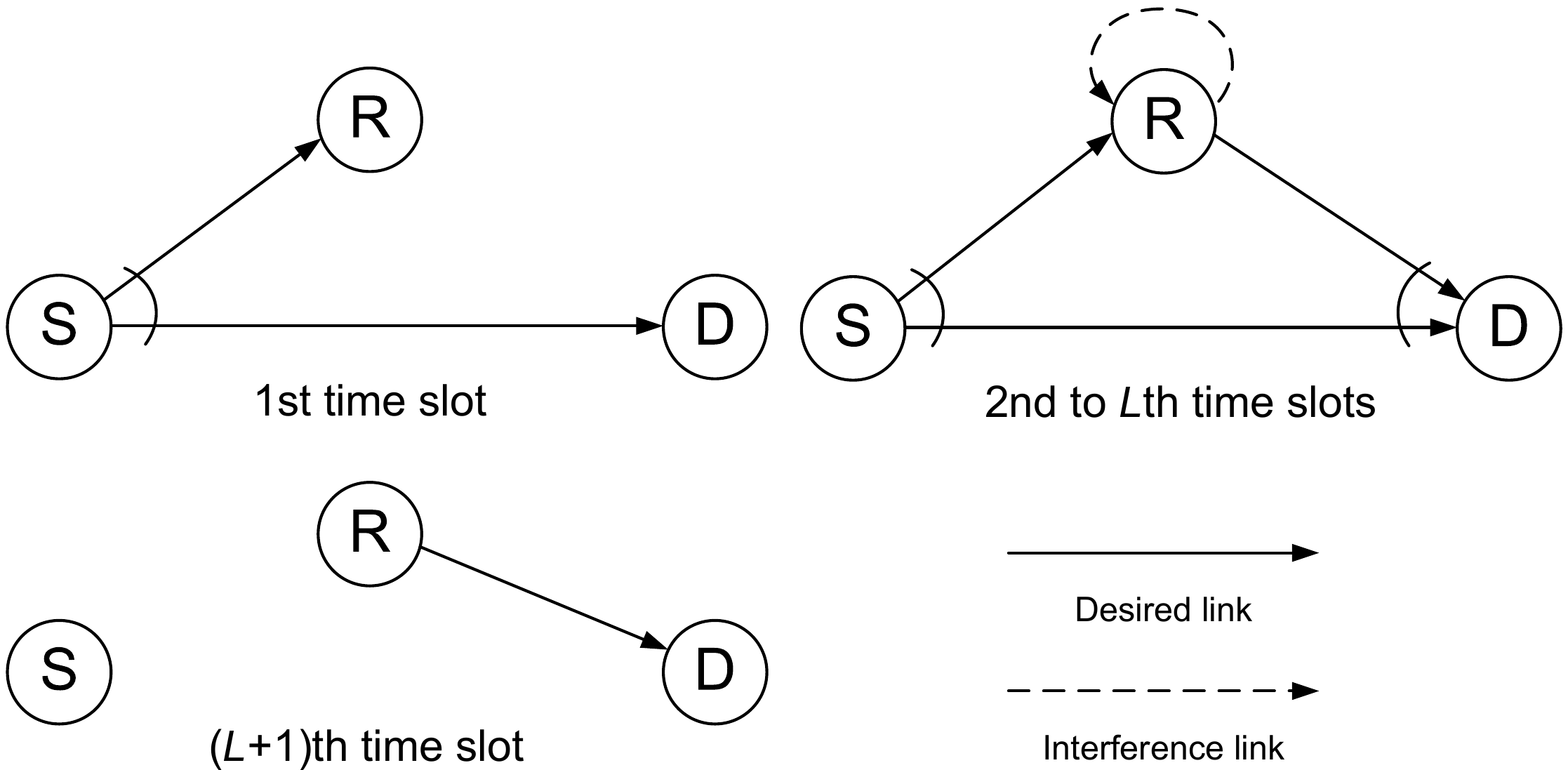,scale=0.45,angle=0}
\end{center}
\caption{Illustration of full-duplex cooperative relaying.}\label{F1}
\end{figure}

%%%%%%%%%%%%%%%%%%%%%%%%%%%%%%%%%%%%%
We assume discrete-time block fading channels, which remain static over each transmission time slot. A three-node relaying network with one source node (S), one destination node (D), and one FD relay node (R) is illustrated in Fig.~\ref{F1}, 
where each node has a single antenna. The source node first encodes the information bits $\mathbf{b}$ using turbo-like channel encoders, e.g., $k$-rate serial concatenated convolutional codes, to generate the coded bits $\mathbf{c}$. Then, $\mathbf{c}$ are mapped into a transmission symbols vector $\mathbf{x}$ based on a $Q$-ary modulation scheme. Subsequently, $\mathbf{x}$ are divided into $L$ frames, and without loss of generality $L$ is assumed to be even. As shown in Fig.~\ref{F1}, the source node broadcasts $L$ frames in $L$ time slots, respectively, and from the second to the $L^{\mathrm{th}}$ time slots, the FD relay node receives and transmits two successive frames per time slot simultaneously. Finally, in the $(L+1)^{\mathrm{th}}$ time slot, the destination node receives the final frame, i.e., the $L^{\mathrm{th}}$ frame sent from the FD relay node.\footnote{In this paper, repetition-coded relaying is assumed for the sake of simplicity, where the FD relay node uses the same encoders and modulation scheme as the source node. For other cases, e.g., distributed turbo code based relaying as in \cite{Valenti2003}, additional diversity gain can be exploited at the cost of decoding complexity at the destination node.} Similar FD relaying procedure can be found in \cite{Khafagy2015}. However, unlike our assumption that channel changes each frame, the authors in \cite{Khafagy2015} assume that channel changes each super-block, where each super-block includes $L$ codewords.
%In the FD mode, R causes self-interference which will affect its simultaneously received signal.  

Denote $\mathbf{y}_{\mathrm{R}}(l)\in\mathcal{C}^{M}$ as the received signal vector in the $l^{\mathrm{th}}$ time slot at the FD relay node, and $\mathbf{y}_{\mathrm{D}}(l)\in\mathcal{C}^{M}$ as the received signal vector in the $l^{\mathrm{th}}$ time slot at the destination node, where $M$ is the number of symbols per frame. Then, we have
\begin{equation}\label{eq01}
\mathbf{y}_{\mathrm{R}}(l)=\sqrt{P_{\mathrm{S}}}h_{\mathrm{S},\mathrm{R}}(l)\mathbf{x}_{\mathrm{S}}(l)+\sqrt{P_{\mathrm{R}}}h_{\mathrm{R},\mathrm{R}}(l)\mathbf{x}_{\mathrm{R}}(l-1)+\mathbf{v}_{\mathrm{R}}(l),
\end{equation}
\begin{equation}\label{eq02}
\mathbf{y}_{\mathrm{D}}(l)=\sqrt{P_{\mathrm{S}}}h_{\mathrm{S},\mathrm{D}}(l)\mathbf{x}_{\mathrm{S}}(l)+\sqrt{P_{\mathrm{R}}}h_{\mathrm{R},\mathrm{D}}(l)\mathbf{x}_{\mathrm{R}}(l-1)+\mathbf{v}_{\mathrm{D}}(l),
\end{equation}
where $h_{\mathrm{S},\mathrm{R}}(l)$, $h_{\mathrm{S},\mathrm{D}}(l)$, and $h_{\mathrm{R},\mathrm{D}}(l)$ are the complex Gaussian channel coefficients with zero mean and variance $\frac{1}{2}\sigma^{2}_{h_{i,j}},i,j\in\{\mathrm{S,R,D}\}$, per dimension in the $l^{\mathrm{th}}$ time slot for S-R link, S-D link, and R-D link, respectively. In addition, $\sigma^{2}_{h_{i,j}}$ follows simplified path loss model with $\sigma^{2}_{h_{i,j}}=d_{i,j}^{-v}$, where $d_{i,j}$ is the distance between node $i$ and node $j$, and $v$ is the path loss exponent; $h_{\mathrm{R},\mathrm{R}}(l)$, subject to the complex Gaussian distribution with zero mean and variance $\frac{1}{2}\sigma^{2}_{h_{R,R}}$ per dimension, is the residual self-interference at the FD relay node in the $l^{\mathrm{th}}$ time slot due to incomplete self-interference cancellation;\footnote{In the literature, many approaches have been proposed to mitigate self-interference effects \cite{Liu2015,Sim2017}. However, self-interference may not be completely cancelled and its residual effect can be modelled as complex Gaussian distributed \cite{Li2017}.} $\mathbf{x}_{\mathrm{S}}(l)\in\mathcal{C}^{M}$ is the $l^{\mathrm{th}}$ frame transmitted in the $l^{\mathrm{th}}$ time slot from the source node; $\mathbf{x}_{\mathrm{R}}(l-1)\in\mathcal{C}^{M}$ is the $(l-1)^{\mathrm{th}}$ frame transmitted in the $l^{\mathrm{th}}$ time slot from the FD relay node, where its formulation method will be introduced in Section III-A; $P_{\mathrm{S}}$ and $P_{\mathrm{R}}$ are the transmit power for the source and the FD relay nodes, respectively; $\mathbf{v}_{\mathrm{R}}(l)\in\mathcal{C}^{M}$ and $\mathbf{v}_{\mathrm{D}}(l)\in\mathcal{C}^{M}$ are the additive white Gaussian noise (AWGN) with zero mean and covariance of $\sigma^{2}_{0}\mathbf{I}_{M}$ for the FD relay and the destination nodes, respectively. Here, $\mathbf{I}_{\mathrm{M}}$ denotes the identity matrix with size of $M$.   

Based on the channel model described above, if the FD relay node can perfectly decode its received signals and has the same transmission power as the source node (i.e., $P\triangleq P_{\mathrm{S}}=P_{\mathrm{R}}$), the entire transmission in $L+1$ time slots from the source node to the destination node with the help of the FD relay node is equivalent to a multiple access multiple-input multiple-output (MIMO) model, which can be expressed as
\begin{equation}\label{eq03}
\mathbf{Y}=\sqrt{P}\mathbf{H}\mathbf{X}+\mathbf{V},
\end{equation}
where 
\begin{equation}
\mathbf{Y}=[\mathbf{y}_{\mathrm{D}}(1),\mathbf{y}_{\mathrm{D}}(2),\ldots,\mathbf{y}_{\mathrm{D}}(L+1)]^{T},
\end{equation}
is the $(L+1)\times M$ received signal matrix at the destination node; 
\begin{equation}
\mathbf{X}=[\mathbf{x}_\mathrm{S}(1),\mathbf{x}_\mathrm{S}(2),\ldots,\mathbf{x}_\mathrm{S}(L)]^{T},
\end{equation}
is the $L\times M$ transmitted signal matrix at the source node; 
\begin{equation}
\mathbf{V}=[\mathbf{v}_{\mathrm{D}}(1),\mathbf{v}_{\mathrm{D}}(2),\ldots,\mathbf{v}_{\mathrm{D}}(L+1)]^{T},
\end{equation}
is the $(L+1)\times M$ AWGN matrix at the destination node; $\mathbf{H}\in\mathcal{C}^{(L+1)\times L}$ is the equivalent channel matrix that is given by
\begin{IEEEeqnarray}{ll}\label{eq04}
\mathbf{H}=\left[\begin{array}{ccccc}
h_{\mathrm{S,D}}(1) & 0 & \cdots & 0 & 0 \\
h_{\mathrm{R,D}}(2) & h_{\mathrm{S,D}}(2) & \cdots & 0 & 0 \\
0 & h_{\mathrm{R,D}}(3) & \cdots & 0 & 0 \\
\vdots & \vdots & \ddots & \vdots & \vdots \\
0 & 0 & \cdots & h_{\mathrm{R,D}}(L) & h_{\mathrm{S,D}}(L) \\
0 & 0 & \cdots & 0 & h_{\mathrm{R,D}}(L+1) \end{array} \right].
\end{IEEEeqnarray}
On the other hand, in practice, the FD relay node may not decode its received frames perfectly due to strong residual self-interference and other channel effects. In this case, an effective method is required to control error propagations.

%Before decoding $L$ frames, D needs to wait $L+1$ tAl-Habian2011,Kwon2eneralized010aime slots until all $L$ frames being received. It then can perform joint decoding to decode all $L$ transmitted frames. 

%%%%%%%%%%%%%%%%%%%%%%%%%%%%%%%%%%%%%%%%%%%%%%%%%%%%%%%%%%%%%%%%%%%%%%%%%%%%%%%%%%%%%%%%%%%%%%%%%%%%%%%%%%%%%%%%

          %%%III. The Proposed Symbol-Level Selective Transmission Scheme %%%

%%%%%%%%%%%%%%%%%%%%%%%%%%%%%%%%%%%%%%%%%%%%%%%%%%%%%%%%%%%%%%%%%%%%%%%%%%%%%%%%%%%%%%%%%%%%%%%%%%%%%%%%%%%%%%%%
\section{The Proposed Symbol-Level Selective Transmission Scheme}
In this section, a square deviation based symbol-level selective relaying scheme is introduced to mitigate error propagation effects and improve system spectral efficiency. The general idea is to predict the positions of the correctly decoded symbols and discard the erroneously decoded ones per frame at the FD relay node, which results in fewer errors being forwarded to the destination node. Then, the modified MAP receiver is implemented at the destination node to mitigate the inter-frame interference and identify the positions of the discarded symbols from the FD relay node for spatial diversity combining.   

%%%%%%%%%%%%%%%%%%%%%%%%%%%%%%%%%%%%%%%%%
\subsection{Symbol-Level Selection Method at the FD Relay}
The proposed symbol-level selection method aims to predict correctly decoded symbols in a frame at the FD relay node based on heuristically calculating the squared Euclidean distance between the reconstructed symbols (i.e. after the successive operations of demodulation/decoding and re-encoding/re-modulation) and the received signal after a linear detection process. Such selection method leads to a low-complexity and high-efficiency symbol-level selective relaying. To elaborate, \eqref{eq01} can be reformulated as a symbol-based equivalent form, which is
\begin{equation}\label{eq05}
y^{(m)}_{\mathrm{R}}(l)=\sqrt{P_{\mathrm{S}}}h_{\mathrm{S},\mathrm{R}}(l)x^{(m)}_{\mathrm{S}}(l)+\sqrt{P_{\mathrm{R}}}h_{\mathrm{R,R}}(l)x^{(m)}_{\mathrm{R}}(l-1)+v^{(m)}_{\mathrm{R}}(l),~\forall m,
\end{equation}
where $y^{(m)}_{\mathrm{R}}(l)$, $x^{(m)}_{\mathrm{S}}(l)$, $x^{(m)}_{\mathrm{R}}(l-1)$ and $v^{(m)}_{\mathrm{R}}(l)$ are the $m^{\mathrm{th}}$ symbol in $\mathbf{y}_{\mathrm{R}}(l)$, $\mathbf{x}_{\mathrm{S}}(l)$, $\mathbf{x}_{\mathrm{R}}(l-1)$ and $\mathbf{v}_{\mathrm{R}}(l)$, respectively. Then, to state the problem as integer square deviation problem and make it work for generalized modulation schemes, we need to find the real-valued equivalent of \eqref{eq05}. To this end, let $\tilde{\mathbf{y}}^{(m)}_{\mathrm{R}}(l)\in\mathcal{R}^{2\times1}$, $\tilde{\mathbf{x}}^{(m)}_{\mathrm{S}}(l)\in\mathcal{R}^{2\times1}$, $\tilde{\mathbf{x}}^{(m)}_{\mathrm{R}}(l-1)\in\mathcal{R}^{2\times1}$ and $\tilde{\mathbf{v}}^{(m)}_{\mathrm{R}}(l)\in\mathcal{R}^{2\times1}$ denote real vectors obtained from $y^{(m)}_{\mathrm{R}}(l)$, $x^{(m)}_{\mathrm{S}}(l)$, $x^{(m)}_{\mathrm{R}}(l-1)$ and $v^{(m)}_{\mathrm{R}}(l)$, respectively, as
\begin{equation}\label{eq06}
\tilde{\mathbf{y}}^{(m)}_{\mathrm{R}}(l)=[\mathcal{R}(y^{(m)}_{\mathrm{R}}(l)),\mathcal{I}(y^{(m)}_{\mathrm{R}}(l))]^{T},
\end{equation}
\begin{equation}\label{eq07}
\tilde{\mathbf{x}}^{(m)}_{\mathrm{S}}(l)=[\mathcal{R}(x^{(m)}_{\mathrm{S}}(l)),\mathcal{I}(x^{(m)}_{\mathrm{S}}(l))]^{T},
\end{equation}
\begin{equation}\label{eq08}
\tilde{\mathbf{x}}^{(m)}_{\mathrm{R}}(l-1)=[\mathcal{R}(x^{(m)}_{\mathrm{R}}(l-1)),\mathcal{I}(x^{(m)}_{\mathrm{R}}(l-1))]^{T}.
\end{equation}
\begin{equation}\label{eq0801}
\tilde{\mathbf{v}}^{(m)}_{\mathrm{R}}(l)=[\mathcal{R}(v^{(m)}_{\mathrm{R}}(l)),\mathcal{I}(v^{(m)}_{\mathrm{R}}(l))]^{T}.
\end{equation}
Additionally, let $\tilde{\mathbf{H}}_{\mathrm{S},\mathrm{R}}(l)\in\mathcal{R}^{2\times2}$ denote real matrix obtained from $\sqrt{P_{\mathrm{S}}}h_{\mathrm{S},\mathrm{R}}(l)$, as
\begin{IEEEeqnarray}{ll}\label{eq09}
\tilde{\mathbf{H}}_{\mathrm{S},\mathrm{R}}(l)=\sqrt{P_{\mathrm{S}}}\left[\begin{array}{cc}
\mathcal{R}(h_{\mathrm{S},\mathrm{R}}(l)) & -\mathcal{I}(h_{\mathrm{S},\mathrm{R}}(l)) \\
\mathcal{I}(h_{\mathrm{S},\mathrm{R}}(l)) & \mathcal{R}(h_{\mathrm{S},\mathrm{R}}(l)) \end{array} \right],
\end{IEEEeqnarray}
and let $\tilde{\mathbf{H}}_{\mathrm{R},\mathrm{R}}(l)\in\mathcal{R}^{2\times2}$ denote real matrix obtained from $\sqrt{P_{\mathrm{R}}}h_{\mathrm{R},\mathrm{R}}(l)$, as
\begin{IEEEeqnarray}{ll}\label{eq0901}
\tilde{\mathbf{H}}_{\mathrm{R},\mathrm{R}}(l)=\sqrt{P_{\mathrm{R}}}\left[\begin{array}{cc}
\mathcal{R}(h_{\mathrm{R},\mathrm{R}}(l)) & -\mathcal{I}(h_{\mathrm{R},\mathrm{R}}(l)) \\
\mathcal{I}(h_{\mathrm{R},\mathrm{R}}(l)) & \mathcal{R}(h_{\mathrm{R},\mathrm{R}}(l)) \end{array} \right].
\end{IEEEeqnarray}
Then, the real-valued equivalent of \eqref{eq05} is given by
\begin{equation}\label{eq10}
\tilde{\mathbf{y}}^{(m)}_{\mathrm{R}}(l)=\tilde{\mathbf{H}}_{\mathrm{S},\mathrm{R}}(l)\tilde{\mathbf{x}}^{(m)}_{\mathrm{S}}(l)+\tilde{\mathbf{H}}_{\mathrm{R},\mathrm{R}}(l)\tilde{\mathbf{x}}^{(m)}_{\mathrm{R}}(l-1)+\tilde{\mathbf{v}}^{(m)}_{\mathrm{R}}(l),~\forall m.
\end{equation}

\begin{figure}[t] 
\begin{center}
\epsfig{figure=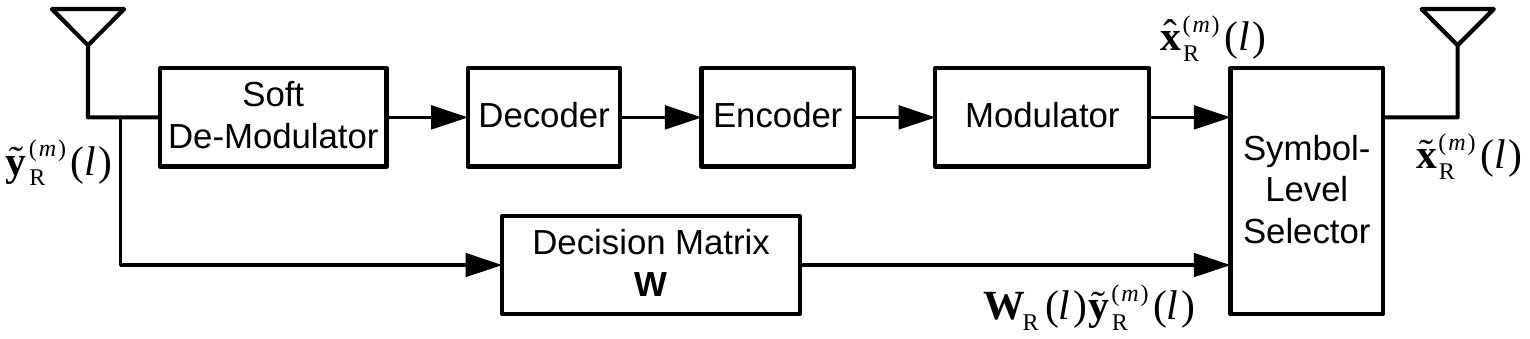,scale=0.7,angle=0}
\end{center}
\caption{Block structure of symbol-level selective relaying.}\label{F2t}
\end{figure}

Fig.~\ref{F2t} shows the basic structure of the proposed symbol-level selective relaying process. 
The received signal $\tilde{\mathbf{y}}^{(m)}_{\mathrm{R}}(l)$ at the FD relay node is first fed into two parallel processing branches: 1) Demodulate/decode and re-encode/re-modulate the potential transmission symbols, i.e., $\hat{\mathbf{x}}^{(m)}_{\mathrm{R}}(l)\in\mathcal{R}^{2\times1},\forall m$, according to the pre-agreed channel coding and modulation methods; 2) Detect the received signal $\tilde{\mathbf{y}}^{(m)}_{\mathrm{R}}(l)$ with a linear detection method. To elaborate, for the first operation, the FD relay node first performs soft demodulation on $\tilde{\mathbf{y}}^{(m)}_{\mathrm{R}}(l),\forall m$ by taking self-interference and AWGN into account, and then the output of demodulator is fed into the turbo-like channel decoders, using the Bahl-Cocke-Jelinek-Raviv (BCJR) algorithm \cite{Bahl1974}, in order to regenerate the original information bits sent from the source node. Then, the FD relay node re-encodes and modulates the information bits according to the pre-agreed encoders/modulator, and converts the complex-valued symbol outputs to the real-valued equivalent symbol vectors, i.e., $\hat{\mathbf{x}}^{(m)}_{\mathrm{R}}(l),\forall m$. For the second operation, let's assume that the instantaneous channel knowledge of the S-R link, the statistical channel knowledge of the R-R link as well as the noise variance are known at the FD relay node. The FD relay node detects $\tilde{\mathbf{y}}^{(m)}_{\mathrm{R}}(l)$ with linear minimum mean-square error (MMSE) detector to minimize MSE between its estimated value and the actual transmitted symbol from the source node. After applying this detection method, channel fading, residual self-interference and noise effects on the received signal can be suppressed simultaneously. Thus, based on \cite{Boyd2009}, this linear MMSE based detection matrix can be formulated as
\begin{equation}\label{eq12}
\mathbf{W}_{\mathrm{R}}(l)=\frac{\sigma^{2}_{x}}{2}\tilde{\mathbf{H}}^{T}_{\mathrm{S},\mathrm{R}}(l)[\frac{\sigma^{2}_{x}}{2}\tilde{\mathbf{H}}_{\mathrm{S},\mathrm{R}}(l)\tilde{\mathbf{H}}^{T}_{\mathrm{S},\mathrm{R}}(l)+\frac{P_{\mathrm{R}}\sigma^{2}_{x}\sigma^{2}_{h_\mathrm{R,R}}}{2}\mathbf{I}_{2}+\frac{\sigma^{2}_{0}}{2}\mathbf{I}_{2}]^{-1}.
\end{equation}
It is worth noting that the decoded bits in the above first operation might not be Gaussian distributed due to the non-linear channel decoding process. However, in our theoretical results, the Gaussian approximation on $\hat{\mathbf{x}}^{(m)}_{\mathrm{R}}(l)$ can still be applied if the FD relay node employs a Gaussian codebook to re-encode/re-modulate the decoded bits, e.g., through turbo-like encoding and $N$-dimensional sphere constellation shaping method \cite{Forney1998}.

%In this paper, the information bits are generated following Gaussian distribution. This is because that the Gaussian distributed signals based on Shannon's theorem can maximize the mutual information between the input and the output of Gaussian channel\cite{Cover2006}. Consequently, the transmitted signals $\tilde{\mathbf{x}}^{(m)}_{\mathrm{S}}(l)$ and $\hat{\mathbf{x}}^{(m)}_{\mathrm{R}}(l)$ can be approximated as Gaussian distributed because the implemented turbo codes in this paper are linear block codes, where the encoding operations can be viewed as the modulo-2 matrix multiplication of the information bits' vector with a generator matrix \cite{Valenti1996}. In addition, modulation shaping may change the distribution of the signal constellation points, but Gaussian assumption can still be applied as an approximation if the $N$-dementional sphere shaping techniques are implemented \cite{Forney1998} .

Following the steps above, the symbol-level selection process can be implemented based on calculating the squared Euclidean distance between the reconstructed symbol and the detected signal with linear MMSE detector, which is
\begin{equation}\label{eq11}
\Delta_m(l)=\|\mathbf{W}_{\mathrm{R}}(l)\tilde{\mathbf{y}}^{(m)}_{\mathrm{R}}(l)-\hat{\mathbf{x}}^{(m)}_{\mathrm{R}}(l)\|^{2},~\forall m.
\end{equation}

Then, to predict the $m^{\mathrm{th}}$ symbol vector $\hat{\mathbf{x}}^{(m)}_{\mathrm{R}}(l)$ as the correctly decoded symbol, we define a utility function as
\begin{IEEEeqnarray}{ll}\label{eq13}
\mathrm{sgn}\left(\varepsilon\right)\triangleq\left\{
\begin{array}{l}
1,~~~~\Delta_m(l)\leq\varepsilon,\\
0,~~~~\mathrm{otherwise},\\
\end{array}
\right.
\end{IEEEeqnarray}
where $\varepsilon$ is the square deviation error threshold. As shown in \eqref{eq13}, if $\mathrm{sgn}\left(\varepsilon\right)=1$, the $m^{\mathrm{th}}$ symbol in the $l^{\mathrm{th}}$ frame is assumed to be decoded correctly at the FD relay node, and be able to forward to the destination node. Otherwise, the transmission power of the symbol will be set to zero as it is predicted as an erroneously decoded symbol. It is worth noting that the symbol $\hat{\mathbf{x}}^{(m)}_{\mathrm{R}}(l)$ in \eqref{eq11} can be considered as a constellation point of the selected modulation scheme. $\varepsilon$ in \eqref{eq13} is used to identify whether the signal $\mathbf{W}_{\mathrm{R}}(l)\tilde{\mathbf{y}}^{(m)}_{\mathrm{R}}(l)$ is the closest to the constellation point $\hat{\mathbf{x}}^{(m)}_{\mathrm{R}}(l)$. Thus, $\varepsilon$ is selected to be square of the half Euclidean distance between two closest constellation points of the selected modulation scheme. With such selection principle of $\epsilon$, it is sufficient to guarantee that $\mathbf{W}_{\mathrm{R}}(l)\tilde{\mathbf{y}}^{(m)}_{\mathrm{R}}(l)$ is the closest to $\hat{\mathbf{x}}^{(m)}_{\mathrm{R}}(l)$ given that their squared Euclidean distance is smaller than $\varepsilon$.
%\footnote{Note that, for some linearly detected signals with high interference/noise effects, there is probability that they are closest to $\hat{\mathbf{x}}^{(m)}_{\mathrm{R}}(l)$ but have been misidentified. However, we use this assumption to simplify the analysis, and we will later demonstrate that there is an agreement between theory and simulations. }  by assuming all the constellation points of a specified modulation scheme with equal vertical and horizontal spacing,

Following the above described procedure, the actual complex-valued transmitted frame at the FD relay node for the next time slot, i.e. $\mathbf{x}_{\mathrm{R}}(l)\in\mathcal{C}^{M}$, can be formulated. In detail, start from formulating the real-valued transmitted frame at the FD relay node for the $(l+1)^{\mathrm{th}}$ time slot, we have
\begin{equation}\label{eq14}
\tilde{\mathbf{x}}^{(m)}_{\mathrm{R}}(l)=\mathrm{sgn}\left(\varepsilon\right)\cdot\hat{\mathbf{x}}^{(m)}_{\mathrm{R}}(l),~\forall m.
\end{equation}
Then, by converting $\tilde{\mathbf{x}}^{(m)}_{\mathrm{R}}(l),\forall m,$ back to the original complex-valued $x^{(m)}_{\mathrm{R}}(l),\forall m$, the actual transmitted frame in complex-valued form at the FD relay node for the $(l+1)^{\mathrm{th}}$ time slot can be obtained by
\begin{equation}\label{eq14t}
\mathbf{x}_{\mathrm{R}}(l)=[x^{(1)}_{\mathrm{R}}(l),x^{(2)}_{\mathrm{R}}(l),\ldots,x^{(M)}_{\mathrm{R}}(l)]^{T}.
\end{equation}

%It is worth noting that, if the $m^{\mathrm{th}}$ symbol in $\mathbf{x}_{\mathrm{R}}(l)$ is not selected, the selection procedure for the $m^{\mathrm{th}}$ symbol in $\mathbf{x}_{\mathrm{R}}(l+1)$ is free from self-interference.

{\em Remark 1:} There is a probability that the reconstructed symbol $\hat{\mathbf{x}}^{(m)}_{\mathrm{R}}(l)$ has been decoded incorrectly while $\Delta_m(l)\leq\varepsilon$. This results in inaccuracy of symbol-level selection. However, this probability is relatively small, especially for high order modulation schemes. The detailed analysis about the accuracy of our proposed symbol-level selection method is provided in Appendix A.

%leads to the most accurate symbol-level selection. However, due to the randomness of instantaneous channels and noise, $\mathbf{W}_{\mathrm{R}}(l)\tilde{\mathbf{y}}^{(m)}_{\mathrm{R}}(l)$ may not always perfectly match with $\hat{\mathbf{x}}^{(m)}_{\mathrm{R}}(l)$. In this case, some correctly decoded symbols may be discarded due to the strict constraint on $\varepsilon_{m}(l)=0$. In Section V, a detailed analysis about $\varepsilon_{m}(l)$ selection is provided, and the result shows that $\varepsilon_{m}(l)=0$ may not give the best performance.  

%%%%%%%%%%%%%%%%%%%%%%%%%%%%%%%%%%%%%%%%%%%%%%%%%%%%%
\subsection{Decoding Process at the Destination}
\begin{figure}[t] 
\begin{center}
\epsfig{figure=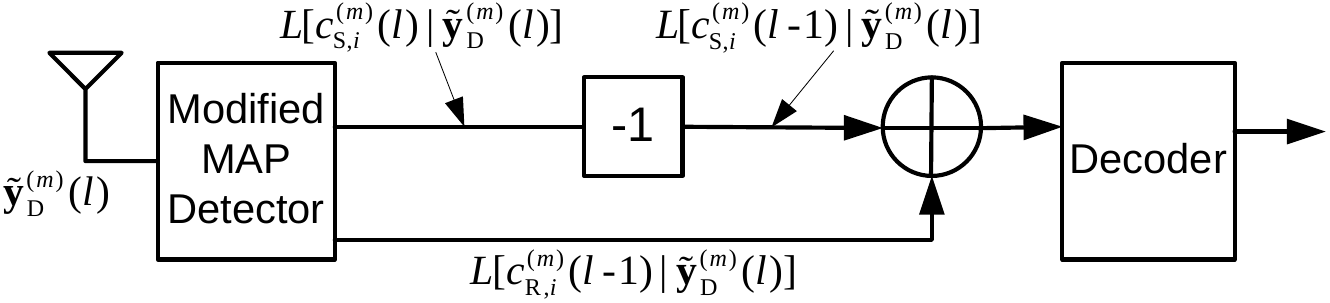,scale=0.7,angle=0}
\end{center}
\caption{Block structure of detection/decoding processes at the destination node.}\label{F2}
\end{figure}

Due to the FD relaying, two successive frames are received simultaneously at the destination node, which results in inter-frame interference. Such interference can be cancelled by the proposed modified MAP detector. In addition, the proposed modified MAP detector can also identify the positions of discarded symbols from the FD relay node for spatial diversity combining. Fig.~\ref{F2} gives the basic structure of detection/decoding processes at the destination node. Here, the ``-1'' box is used to shift the frame index, so that we can combine two versions of LLR values per frame sent from both source and FD relay nodes.

For the purpose of diversity combining and turbo-like decoding, our proposed modified MAP detector should provide the soft information, i.e., the LLR values, at its output. To this end, let's assume the destination node at the current $l^{\mathrm{th}}$ time instant expects to decode the symbols $x^{(m)}_{\mathrm{S}}(l)$ and $x^{(m)}_{\mathrm{R}}(l-1)$ simultaneously. In this case, we first construct $\tilde{\mathbf{x}}\triangleq[\tilde{x}^{(m)}_{\mathrm{S},1}(l),\tilde{x}^{(m)}_{\mathrm{R},1}(l-1),\tilde{x}^{(m)}_{\mathrm{S},2}(l),\tilde{x}^{(m)}_{\mathrm{R},2}(l-1)]$ as their trial vector, where $\tilde{x}^{(m)}_{\mathrm{S},1}(l)$ and $\tilde{x}^{(m)}_{\mathrm{R},1}(l-1)$ are the trial elements for the real parts of $x^{(m)}_{\mathrm{S}}(l)$ and $x^{(m)}_{\mathrm{R}}(l-1)$, respectively; $\tilde{x}^{(m)}_{\mathrm{S},2}(l)$ and $\tilde{x}^{(m)}_{\mathrm{R},2}(l-1)$ are the trial elements for the imaginary parts of $x^{(m)}_{\mathrm{S}}(l)$ and $x^{(m)}_{\mathrm{R}}(l-1)$, respectively. Thus, the received signal at the destination node in real-valued form can be expressed as
\begin{equation}\label{eqAD02m}
\tilde{\mathbf{y}}^{(m)}_{\mathrm{D}}(l)=\tilde{\mathbf{H}}_{\mathrm{D}}(l)\tilde{\mathbf{x}}^{T}+\tilde{\mathbf{v}}^{(m)}_{D}(l),
\end{equation}
where 
\begin{equation}\label{eqAD02}
\tilde{\mathbf{y}}^{(m)}_{\mathrm{D}}(l)\triangleq[\mathcal{R}(y^{(m)}_{\mathrm{D}}(l)),\mathcal{I}(y^{(m)}_{\mathrm{D}}(l))]^{T},
\end{equation}
and $y^{(m)}_{\mathrm{D}}(l)$ in \eqref{eqAD02} is the $m^{\mathrm{th}}$ symbol of $\mathbf{y}_{\mathrm{D}}(l)$; 
\begin{equation}\label{eqAD02tt}
\tilde{\mathbf{v}}^{(m)}_{\mathrm{D}}(l)\triangleq[\mathcal{R}(v^{(m)}_{\mathrm{D}}(l)),\mathcal{I}(v^{(m)}_{\mathrm{D}}(l))]^{T},
\end{equation}
and $v^{(m)}_{\mathrm{D}}(l)$ in \eqref{eqAD02tt} is the $m^{\mathrm{th}}$ symbol of $\mathbf{v}_{\mathrm{D}}(l)$; In addition,
\begin{IEEEeqnarray}{ll}\label{eqAD03}
\tilde{\mathbf{H}}_{\mathrm{D}}(l)\triangleq\left[\begin{array}{cc}
\mathcal{R}(\mathbf{h}_{\mathrm{D}}(l)) & -\mathcal{I}(\mathbf{h}_{\mathrm{D}}(l)) \\
\mathcal{I}(\mathbf{h}_{\mathrm{D}}(l)) & \mathcal{R}(\mathbf{h}_{\mathrm{D}}(l)) \end{array} \right], 
\end{IEEEeqnarray}
and $\mathbf{h}_{\mathrm{D}}(l)=[\sqrt{P_{\mathrm{S}}}h_{\mathrm{S,D}}(l),\sqrt{P_{\mathrm{R}}}h_{\mathrm{R,D}}(l)]$ following \eqref{eq02} in Sec. II. On the other hand, by considering $Q$-ary complex modulation scheme, each modulated symbol can be constructed by $Z\triangleq\log_{2} Q$ coded bits, i.e., $x^{(m)}_{\mathrm{S}}(l)$ is constructed by $[c^{(m)}_{\mathrm{S},1}(l),c^{(m)}_{\mathrm{S},2}(l),\ldots,c^{(m)}_{\mathrm{S},Z}(l)]$ and $x^{(m)}_{\mathrm{R}}(l-1)$ is constructed by $[c^{(m)}_{\mathrm{R},1}(l-1),c^{(m)}_{\mathrm{R},2}(l-1),\ldots,c^{(m)}_{\mathrm{R},Z}(l-1)]$. Then, following the conventional MAP detector design as in \cite{Hochwald2003,Vikalo2004}, the LLR value generated from our modified MAP detector for the $i^{\mathrm{th}}$ coded bit of the $m^{\mathrm{th}}$ symbol sent from the source node can be formulated as
\begin{eqnarray}\label{eqAD01}
L[c^{(m)}_{\mathrm{S},i}(l)|\tilde{\mathbf{y}}^{(m)}_{\mathrm{D}}(l)]&=&\log\frac{\mathrm{Pr}[c^{(m)}_{\mathrm{S},i}(l)=0|\tilde{\mathbf{y}}^{(m)}_{\mathrm{D}}(l)]}{\mathrm{Pr}[c^{(m)}_{\mathrm{S},i}(l)=1|\tilde{\mathbf{y}}^{(m)}_{\mathrm{D}}(l)]}\nonumber\\
&\overset{(a)}{=}& \log\frac{\sum_{\tilde{\mathbf{x}}:c^{(m)}_{\mathrm{S},i}(l)=0}\mathrm{Pr}[\tilde{\mathbf{y}}^{(m)}_{\mathrm{D}}(l)|\tilde{\mathbf{x}}]\prod_{j}\mathrm{Pr}[\tilde{x}_{j}]}{\sum_{\tilde{\mathbf{x}}:c^{(m)}_{\mathrm{S},i}(l)=1}\mathrm{Pr}[\tilde{\mathbf{y}}^{(m)}_{\mathrm{D}}(l)|\tilde{\mathbf{x}}]\prod_{j}\mathrm{Pr}[\tilde{x}_{j}]}\nonumber\\
&\overset{(b)}{=}&\log\frac{\sum_{\tilde{\mathbf{x}}:c^{(m)}_{\mathrm{S},i}(l)=0}e^{-\|\tilde{\mathbf{y}}^{(m)}_{\mathrm{D}}(l)-\tilde{\mathbf{H}}_{\mathrm{D}}(l)\tilde{\mathbf{x}}^{T}\|^2}}{\sum_{\tilde{\mathbf{x}}:c^{(m)}_{\mathrm{S},i}(l)=1}e^{-\|\tilde{\mathbf{y}}^{(m)}_{\mathrm{D}}(l)-\tilde{\mathbf{H}}_{\mathrm{D}}(l)\tilde{\mathbf{x}}^{T}\|^2}},~\forall i,
\end{eqnarray}
where (a) is obtained by following Bayes' rule and assuming the elements in $\tilde{\mathbf{x}}$ (i.e., $\tilde{x}_{j},\forall j$) are independent; To obtain (b), assume that the instantaneous channel matrix $\tilde{\mathbf{H}}_{\mathrm{D}}(l)$ in \eqref{eqAD02m} is known at the destination node, and the noise $\tilde{\mathbf{v}}^{(m)}_{D}(l)$ follows Gaussian distribution with zero mean and unit variance per element. Then, the conditional probability of $\tilde{\mathbf{y}}^{(m)}_{\mathrm{D}}(l)$ given $\tilde{\mathbf{x}}$ is
\begin{equation}\label{eqADtony}
\mathrm{Pr}[\tilde{\mathbf{y}}^{(m)}_{\mathrm{D}}(l)|\tilde{\mathbf{x}}]=\frac{1}{(2\pi)^2}e^{-\|\tilde{\mathbf{y}}^{(m)}_{\mathrm{D}}(l)-\tilde{\mathbf{H}}_{\mathrm{D}}(l)\tilde{\mathbf{x}}^{T}\|^2}.
\end{equation}
By inserting \eqref{eqADtony} into (a), we obtain (b). Here, $\prod_{j}\mathrm{Pr}[\tilde{x}_{j}]$ in (a) can be removed from (b) since we assume that all symbols are equiprobable at the initial stage of decoding process. It is worth noting that the trial element combinations of $\tilde{\mathbf{x}}$ in \eqref{eqAD01} for our modified MAP detector should take $\tilde{x}^{(m)}_{\mathrm{R},j}(l-1)=0,\forall j$, into account, where $\tilde{x}^{(m)}_{\mathrm{R},j}(l-1)=0,\forall j$, denotes the corresponding symbol that is discarded at the FD relay node. 

On the other hand, in order to calculate the LLR value for the $i^{\mathrm{th}}$ coded bit of the $m^{\mathrm{th}}$ symbol sent from the FD relay node and identify whether it has been discarded, we first calculate $\mathrm{Pr}[c^{(m)}_{\mathrm{R},i}(l-1)=0|\tilde{\mathbf{y}}^{(m)}_{\mathrm{D}}(l)]$, $\mathrm{Pr}[c^{(m)}_{\mathrm{R},i}(l-1)=1|\tilde{\mathbf{y}}^{(m)}_{\mathrm{D}}(l)]$, and $\mathrm{Pr}[c^{(m)}_{\mathrm{R},i}(l-1)=\varnothing|\tilde{\mathbf{y}}^{(m)}_{\mathrm{D}}(l)]$, $\forall i$ according to Bayes' rule and the assumptions discussed in last paragraph. Here, $c^{(m)}_{\mathrm{R},i}(l-1)=\varnothing$ in $\mathrm{Pr}[c^{(m)}_{\mathrm{R},i}(l-1)=\varnothing|\tilde{\mathbf{y}}^{(m)}_{\mathrm{D}}(l)]$ represents the coded bits of the $m^{\mathrm{th}}$ symbol is discarded at the FD relay node. Then, the LLR value for the $i^{\mathrm{th}}$ coded bit of the $m^{\mathrm{th}}$ symbol sent from the FD relay node can be calculated by considering the following \textit{If} conditions:\\ 
$\ast~$\textit{If} the largest probability among $\mathrm{Pr}[c^{(m)}_{\mathrm{R},i}(l-1)=0|\tilde{\mathbf{y}}^{(m)}_{\mathrm{D}}(l)],\forall i,$ and $\mathrm{Pr}[c^{(m)}_{\mathrm{R},i}(l-1)=1|\tilde{\mathbf{y}}^{(m)}_{\mathrm{D}}(l)],\forall i$, is smaller than the smallest probability among $\mathrm{Pr}[c^{(m)}_{\mathrm{R},i}(l-1)=\varnothing|\tilde{\mathbf{y}}^{(m)}_{\mathrm{D}}(l)],\forall i$, we have
\begin{equation}\label{eqAD05}
L[c^{(m)}_{\mathrm{R},i}(l-1)|\tilde{\mathbf{y}}^{(m)}_{\mathrm{D}}(l)]=0,~\forall i.
\end{equation}
$\ast~~$\textit{Otherwise},
\begin{eqnarray}\label{eqAD04}
L[c^{(m)}_{\mathrm{R},i}(l-1)|\tilde{\mathbf{y}}^{(m)}_{\mathrm{D}}(l)]=\log\frac{\sum_{\tilde{\mathbf{x}}:c^{(m)}_{\mathrm{R},i}(l-1)=0}e^{-\|\tilde{\mathbf{y}}^{(m)}_{\mathrm{D}}(l)-\tilde{\mathbf{H}}_{\mathrm{D}}(l)\tilde{\mathbf{x}}^{T}\|^2}}{\sum_{\tilde{\mathbf{x}}:c^{(m)}_{\mathrm{R},i}(l-1)=1}e^{-\|\tilde{\mathbf{y}}^{(m)}_{\mathrm{D}}(l)-\tilde{\mathbf{H}}_{\mathrm{D}}(l)\tilde{\mathbf{x}}^{T}\|^2}},~\forall
i.
\end{eqnarray}
The zero LLR value in \eqref{eqAD05} denotes the symbol has been discarded at the FD relay node, and the LLR value in \eqref{eqAD04} follows the same derivation procedure of \eqref{eqAD01}. It is worth noting that the trial element combinations in vector $\tilde{\mathbf{x}}$ of \eqref{eqAD04} should not consider the case that $\tilde{x}^{(m)}_{\mathrm{R},j}(l-1)=0,\forall j$ as \eqref{eqAD01}, since \eqref{eqAD05} has already taken the discarded symbol into account. 

Based on \eqref{eqAD01}, \eqref{eqAD05} and \eqref{eqAD04}, the destination node is able to cancel the inter-frame interference and identify the position of discarded symbols from the FD relay node. Subsequently, it needs to wait all the frames being detected and then combine the LLR values of the frame sent from the source node in the previous time slot with the corresponding LLR values of the frame sent from the FD relay node in current time slot for the turbo-like decoding process. Different from the conventional methods in \cite{Al-Habian2011,Kwon2010a,Yi2015}, our proposed scheme omits the feed forwarding step at the FD relay node and avoids the additional signalling overhead.

%%%%%%%%%%%%%%%%%%%%%%%%%%%%%%%%%%%%%%%%%%%%%%%%%%%%%%%%%%%%%%%%%%%%%%%%%%%%%%%%%%%%%%%%%%%%%%%%%%%%%%%%%%%%%%%%%%%%%%%%%%%%%%%%%%%%%%%%%%%%%%%%%%%%%%%%

          %%%VI. Outage Probabilities Analysis%%%

%%%%%%%%%%%%%%%%%%%%%%%%%%%%%%%%%%%%%%%%%%%%%%%%%%%%%%%%%%%%%%%%%%%%%%%%%%%%%%%%%%%%%%%%%%%%%%%%%%%%%%%%%%%%%%%%%%%%%%%%%%%%%%%%%%%%%%%%%%%%%%%%%%%%%%%%
\section{Outage Probability based Power and Location Optimization}
In this section, we first analyse outage probability of our proposed symbol-level selective FD relaying scheme. Then, we find the optimal power allocation and relay location placement so that the outage probability is minimized. In addition, we also provide the outage analysis of HD based symbol-level selection scheme for comparison.

%Here, if a symbol is discarded in the previous time slot, there will be no self-interference effect for the corresponding symbol in the current time slot. We should take this factor into account when we calculate $\mathcal{P}_{\mathrm{C}}$.we start from the case that the selection decision made for the current symbol including self-interference effect. 
%%%%%%%%%%%%%%%%%%%%%%%%%%%%%%%%%%%%%%%%%%%%%%%%%%%%
\subsection{Outage Probability Analysis}
In order to exploit the outage performance of our proposed scheme, the average probability of correctly predicted/forwarded symbols per frame (i.e. $\mathcal{P}_{\mathrm{C}}$) at the FD relay node first needs to be obtained. Specifically, assume that $\tilde{\mathbf{H}}_{\mathrm{S},\mathrm{R}}(l)$ in \eqref{eq10} is known and fixed for the $l^{\mathrm{th}}$ time slot; $\tilde{\mathbf{H}}_{\mathrm{R},\mathrm{R}}(l)$, $\hat{\mathbf{x}}^{(m)}_{\mathrm{R}}(l)$ and $\tilde{\mathbf{v}}^{(m)}_{\mathrm{R}}(l)$ follow Gaussian distribution with zero mean and variances of $\sigma^{2}_{h_{\mathrm{R,R}}}/2$, $\sigma^{2}_{x}/2$ and $\sigma^{2}_{0}/2$ per element, respectively. Then, according to the work in \cite{Boyd2009}, the deviation vector $\mathbf{W}_{\mathrm{R}j}(l)\tilde{\mathbf{y}}^{(m)}_{\mathrm{R}j}(l)-\hat{\mathbf{x}}^{(m)}_{\mathrm{R}}(l)$ follows Gaussian distribution with zero mean and covariance matrix as
\begin{IEEEeqnarray}{ll}\label{eq15}
\mathbf{C}_{e}\triangleq\frac{\sigma^{2}_{x}}{2}\mathbf{I}_2-\frac{\sigma^{4}_{x}}{4}\tilde{\mathbf{H}}^{T}_{\mathrm{S},\mathrm{R}}(l)[\frac{\sigma^{2}_{x}}{2}\tilde{\mathbf{H}}_{\mathrm{S},\mathrm{R}}(l)\tilde{\mathbf{H}}^{T}_{\mathrm{S},\mathrm{R}}(l)+\frac{P_{\mathrm{R}}\sigma^{2}_{h_{\mathrm{R,R}}}\sigma^{2}_{x}}{2}\mathbf{I}_2+\frac{\sigma^{2}_{0}}{2}\mathbf{I}_2]^{-1}\tilde{\mathbf{H}}_{\mathrm{S},\mathrm{R}}(l),
\end{IEEEeqnarray}
where $\tilde{\mathbf{H}}_{\mathrm{S},\mathrm{R}}(l)\tilde{\mathbf{H}}^{T}_{\mathrm{S},\mathrm{R}}(l)=P_{\mathrm{S}}\tilde{\sigma}^{2}_{h_{\mathrm{S,R}}}\mathbf{I}_2$ is a scalar matrix, and $\tilde{\sigma}^{2}_{h_{\mathrm{S,R}}}$ is the instantaneous channel gain of S-R link. Then, \eqref{eq15} can be further simplified to 
\begin{IEEEeqnarray}{ll}\label{eq16}
\mathbf{C}_{e}=\underbrace{\left(\frac{\sigma^{2}_{x}}{2}-\frac{P_{\mathrm{S}}\sigma^{4}_{x}\tilde{\sigma}^{2}_{h_{\mathrm{S,R}}}}{2P_{\mathrm{S}}\sigma^{2}_{x}\tilde{\sigma}^{2}_{h_{\mathrm{S,R}}}+2P_{\mathrm{R}}\sigma^{2}_{x}\sigma^{2}_{h_{\mathrm{R,R}}}+2\sigma^{2}_{0}}\right)}_{\triangleq\sigma^{2}_{C_e}}\mathbf{I}_2,
\end{IEEEeqnarray}
which is also a scalar matrix with the scalar value defined as $\sigma^{2}_{C_e}$. Thus, the square deviation value $\Delta_m(l)$ follows the chi-squared distribution with two degrees of freedom. In this case, the probability of the $m^{\mathrm{th}}$ symbol in the $l^{\mathrm{th}}$ frame that is selected to be forwarded to the destination node can be given by
\begin{eqnarray}\label{eq17}
\mathcal{P}_{\mathrm{S}}&\triangleq&\mathrm{Pr}(\Delta_m(l)\leq\varepsilon),\nonumber\\
&=&\int^{\frac{\varepsilon}{\sigma^{2}_{C_e}}}_{0}f_{\Delta}(x;2)dx,\nonumber\\
&=&1-e^{-\frac{\varepsilon}{2\sigma^{2}_{C_e}}},~\forall m,l,
\end{eqnarray}
where $f_{\Delta}(x;2)$ in \eqref{eq17} is the probability density function (p.d.f.) of $\Delta_m(l)$. It is worth noting that, due to statistical identity, all the symbols are with the same $\mathcal{P}_{\mathrm{S}}$. In addition, if the $m^{\mathrm{th}}$ symbol of a frame is discarded in the $l^{\mathrm{th}}$ time slot, there will be no self-interference effect for the $m^{\mathrm{th}}$ symbol of a frame in the next time slot. In this case, we should set the term $2P_{\mathrm{R}}\sigma^{2}_{x}\sigma^{2}_{h_{\mathrm{R,R}}}$ in \eqref{eq16} to be zero.

\begin{figure}[t] 
\begin{center}
\epsfig{figure=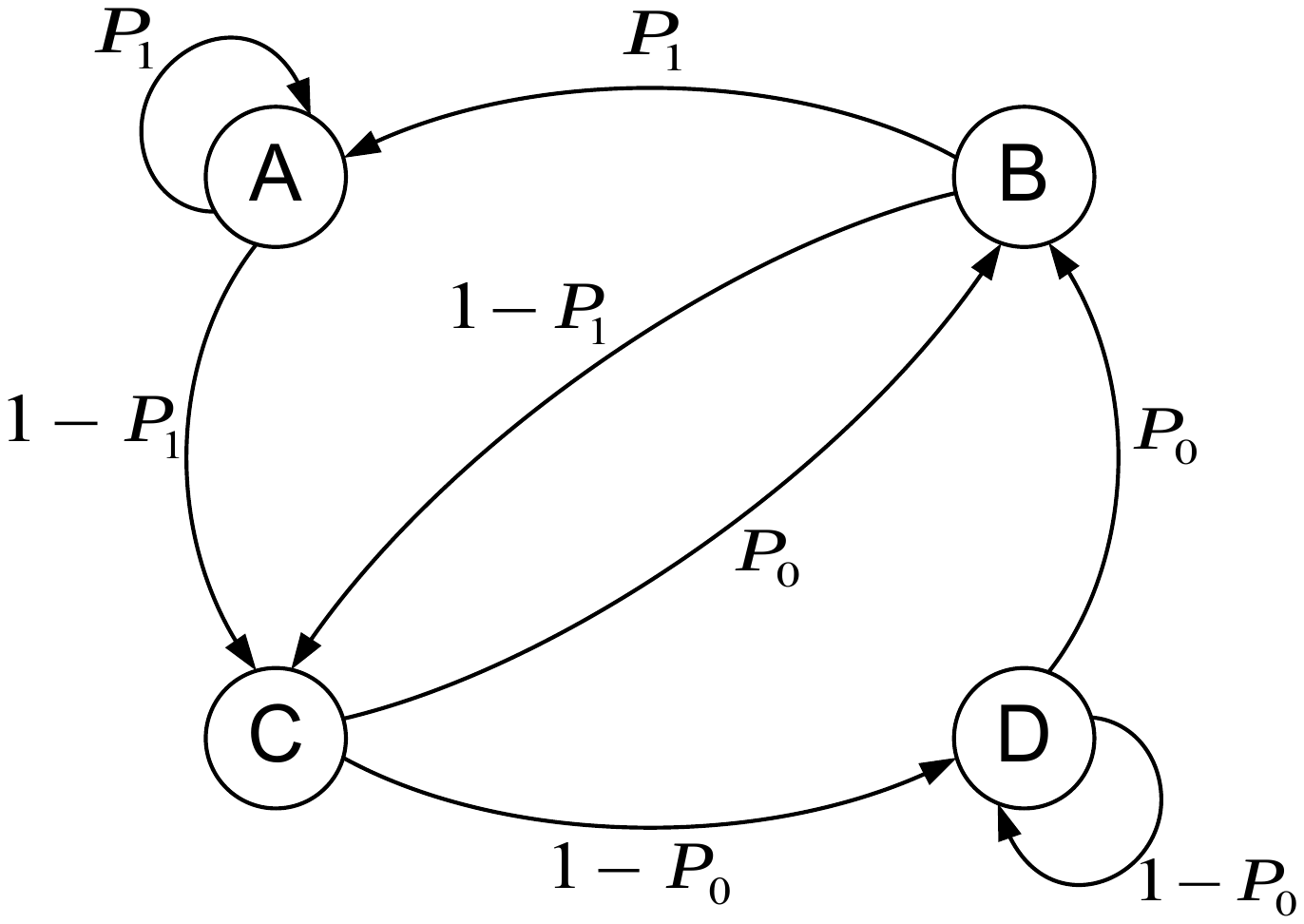,scale=0.6,angle=0}
\end{center}
\caption{The state transition diagram of Markov chain for calculating the average probability of correctly predicted/forwarded symbols per frame.
}\label{F3a}
\end{figure}

Since $\mathcal{P}_{\mathrm{S}}$ is based solely on the selection decision from its previous time slot,  then, the average probability $\mathcal{P}_{\mathrm{C}}$ can be calculated with the help of Markov chain modeling \cite{Doob1953}. To elaborate, four states should be considered: A) a symbol is selected on the condition that self-interference is present; B) a symbol is selected on the condition that self-interference is not present; C) a symbol is not selected on the condition that self-interference is present; D) a symbol is not selected on the condition that self-interference is not present. Then, we can formulate the state transition diagram as shown in Fig.~\ref{F3a}, and its corresponding state transition matrix is given by 
\begin{IEEEeqnarray}{ll}\label{eq18}
\mathbf{T}=
\left[\begin{array}{cccc}
\mathcal{P}_{1} & 0 & 1-\mathcal{P}_{1} & 0 \\
\mathcal{P}_{1} & 0 & 1-\mathcal{P}_{1} & 0 \\
0 & \mathcal{P}_{0} & 0 & 1-\mathcal{P}_{0} \\
0 & \mathcal{P}_{0} & 0 & 1-\mathcal{P}_{0} \end{array} \right],
\end{IEEEeqnarray}
where $\mathcal{P}_{1}$ is the probability of \eqref{eq17} on the condition that the term $2P_{\mathrm{R}}\sigma^{2}_{x}\sigma^{2}_{h_{\mathrm{R,R}}}$ in \eqref{eq16} is nonzero, and $\mathcal{P}_{0}$ is the probability of \eqref{eq17} on the condition that the term $2P_{\mathrm{R}}\sigma^{2}_{x}\sigma^{2}_{h_{\mathrm{R,R}}}$ in \eqref{eq16} is equal to zero. Then, $\mathcal{P}_{\mathrm{C}}$ can be calculated by
\begin{equation}\label{eq19}
\mathcal{P}_{\mathrm{C}}=\frac{1}{LM}\sum^{M}_{m=1}\sum^{L}_{l=1}\mathbf{u}\mathbf{T}^{l-1}\mathbf{v},
\end{equation}
where $\mathbf{u}\triangleq[0,\mathcal{P}_{0},0,1-\mathcal{P}_0]$ is used to formulate the starting state of the Markov chain. In this case, we predict the correctly decoded symbols of the first frame, where the self-interference should not be present (i.e. B or D). In addition, $\mathbf{v}\triangleq[1,1,0,0]^{\mathrm{T}}$ is used to sum up the probability events that the symbol is selected in a time slot (i.e. A and B).

Since the symbol-level selection method includes the forwarding and the non-forwarding modes, the overall outage probability is the sum of the products of each mode's occurrence probability and the outage probability, which can be expressed as
\begin{equation}\label{eq20}
\mathcal{P}^{\mathrm{FD}}_{\mathrm{out}}=\mathcal{P}_{\mathrm{C}}\mathcal{P}^{\mathrm{FD}}_{\mathrm{FW}}+(1-\mathcal{P}_{\mathrm{C}})\mathcal{P}^{\mathrm{FD}}_{\mathrm{Non-FW}},
\end{equation}
where $\mathcal{P}^{\mathrm{FD}}_{\mathrm{FW}}$ denotes system outage probability when the symbol at the FD relay node is predicted to be correctly decoded and forwarded, otherwise the probability is $\mathcal{P}^{\mathrm{FD}}_{\mathrm{Non-FW}}$. Then, we have the following proposition.

\textit{Proposition 1:} By defining $X\triangleq P_{\mathrm{S}}\sigma^{2}_{h_{\mathrm{S,D}}}$ and  $Y\triangleq P_{\mathrm{R}}\sigma^{2}_{h_{\mathrm{R,D}}}$, and letting $\sigma^{2}_{x}=1$, the overall system outage probability can be expressed as 
\begin{IEEEeqnarray}{ll}\label{eq21}
\mathcal{P}^{\mathrm{FD}}_{\mathrm{out}}=\left\{
\begin{array}{l}
\mathcal{P}_{\mathrm{C}}\left(1-\frac{e^{R}-1+X}{X}e^{-\frac{e^{R}-1}{X}}\right)+(1-\mathcal{P}_{\mathrm{C}})\left(1-e^{-\frac{e^{R}-1}{X}}\right),~~~~~~~~~~~~~~~~~~X=Y,\\
\mathcal{P}_{\mathrm{C}}\bigg( 1-\frac{Y}{Y-X}e^{-\frac{e^{R}-1}{Y}}+\frac{X}{Y-X}e^{-\frac{e^{R}-1}{X}}\bigg)+(1-\mathcal{P}_{\mathrm{C}})\left(1-e^{-\frac{e^{R}-1}{X}}\right),~~~~X\neq Y,\\
\end{array}
\right.
\end{IEEEeqnarray}
where $R$ is the system target transmission rate and $\mathcal{P}_{\mathrm{C}}$ is formulated from \eqref{eq19}.
\begin{proof}
See Appendix B.
\end{proof}

%%%%%%%%%%%%%%%%%%%%%%%%%%%%%%%%%%%%%%%%%%%%%%%%%%%%
\subsection{Power and Location Optimization}
Using the derived system outage probability in \eqref{eq21}, the optimal transmission power allocation and relay location placement can be obtained by solving the following problem, which is
\begin{IEEEeqnarray}{ll}\label{eq23}
\underset{P_{\mathrm{S}},P_{\mathrm{R}},d_{\mathrm{S,R}},d_{\mathrm{R,D}}}{\mathrm{minimize}}&~~\mathcal{P}^{\mathrm{FD}}_{\mathrm{out}}(P_\mathrm{S},P_{\mathrm{R}},d_{\mathrm{S,R}},d_{\mathrm{R,D}})\\
\mathrm{~~subject~to}&~~d_{\mathrm{S,R}}+d_{\mathrm{R,D}}=d_{\mathrm{S,D}},~d_{\mathrm{S,R}}\geq0,d_{\mathrm{R,D}}\geq0,\nonumber\\
&~~P_{\mathrm{S}}+P_{\mathrm{R}}\leq P_{\mathrm{tot}},~P_{\mathrm{S}}\geq0,P_{\mathrm{R}}\geq0;\nonumber
\end{IEEEeqnarray}
where $d_{\mathrm{S,R}}+d_{\mathrm{R,D}}=d_{\mathrm{S,D}}$ denotes the FD relay node is placed on the straight line between the source node and the destination node, and $P_{\mathrm{tot}}$ is a joint total power constraint for the source node and the FD relay node.\footnote{The power optimization based on the total power constraint provides useful insight into the power usage of the whole system. In addition, the relay location placement constraint can be extended to other cases as long as there is a unique relation among the three nodes.} Since the specific expression of outage probability depends on the relation between $X$ and $Y$ in \eqref{eq21}, we will analyse both cases and find their optimal power and location solutions.%\footnote{Due to the sum power and relay location constraints, once the optimal $P_{\mathrm{S}}$ and $d_{\mathrm{S,R}}$ are found, the optimal $P_{\mathrm{R}}$ and $d_{\mathrm{R,D}}$ can be found as well.} 

Specifically, consider the case where $X=Y$ (i.e. $\frac{P_{\mathrm{S}}}{d^{2}_{\mathrm{S,D}}}=\frac{P_{\mathrm{R}}}{d^{2}_{\mathrm{R,D}}}$). By taking  $P_{\mathrm{tot}}=P_{\mathrm{S}}+P_{\mathrm{R}}$ into account, we have 
\begin{equation}\label{eq24}
P_{\mathrm{S}}=\frac{P_{\mathrm{tot}}d^{2}_{\mathrm{S,D}}}{(d_{\mathrm{S,D}}-d_{\mathrm{S,R}})^2+d^{2}_{\mathrm{S,D}}}.
\end{equation}
Then, by inserting \eqref{eq24} into the outage probability expression and replacing $P_{\mathrm{R}}$ with $P_{\mathrm{tot}}-P_{\mathrm{S}}$, we can formulate the outage function, i.e., $\mathcal{P}^{\mathrm{FD}}_{\mathrm{out}}(d_{\mathrm{S,R}})$, with $d_{\mathrm{S,R}}$ as its unique random variable. In this case, with some mathematical manipulations, we can prove that the second order derivative of $\mathcal{P}^{\mathrm{FD}}_{\mathrm{out}}(d_{\mathrm{S,R}})$ with respect to (w.r.t.) $d_{\mathrm{S,R}}\in[0,d_{\mathrm{S,D}}]$ is larger than zero, i.e., $\frac{\partial^{2}\mathcal{P}^{\mathrm{FD}}_{\mathrm{out}}(d_{\mathrm{S,R}})}{\partial d^{2}_{\mathrm{S,R}}}>0$, hence the optimal solution can be obtained by finding the unique root of equation $\frac{\partial\mathcal{P}^{\mathrm{FD}}_{\mathrm{out}}(d_{\mathrm{S,R}})}{\partial d_{\mathrm{S,R}}}=0$ within the interval $[0,d_{\mathrm{S,D}}]$. However, since $\mathcal{P}^{\mathrm{FD}}_{\mathrm{out}}(d_{\mathrm{S,R}})$ is a fairly complicated function, the closed-form expression of the optimal $d_{\mathrm{S,R}}$ which lead to the minimum $\mathcal{P}^{\mathrm{FD}}_{\mathrm{out}}(d_{\mathrm{S,R}})$ is not easy to find. Thus, in this paper, we resort to the numerical bisection search method \cite{Burden2011}, whereby the optimal solution can be found with around fifteen iterations.  

\begin{figure}[t] 
\begin{center}
\epsfig{figure=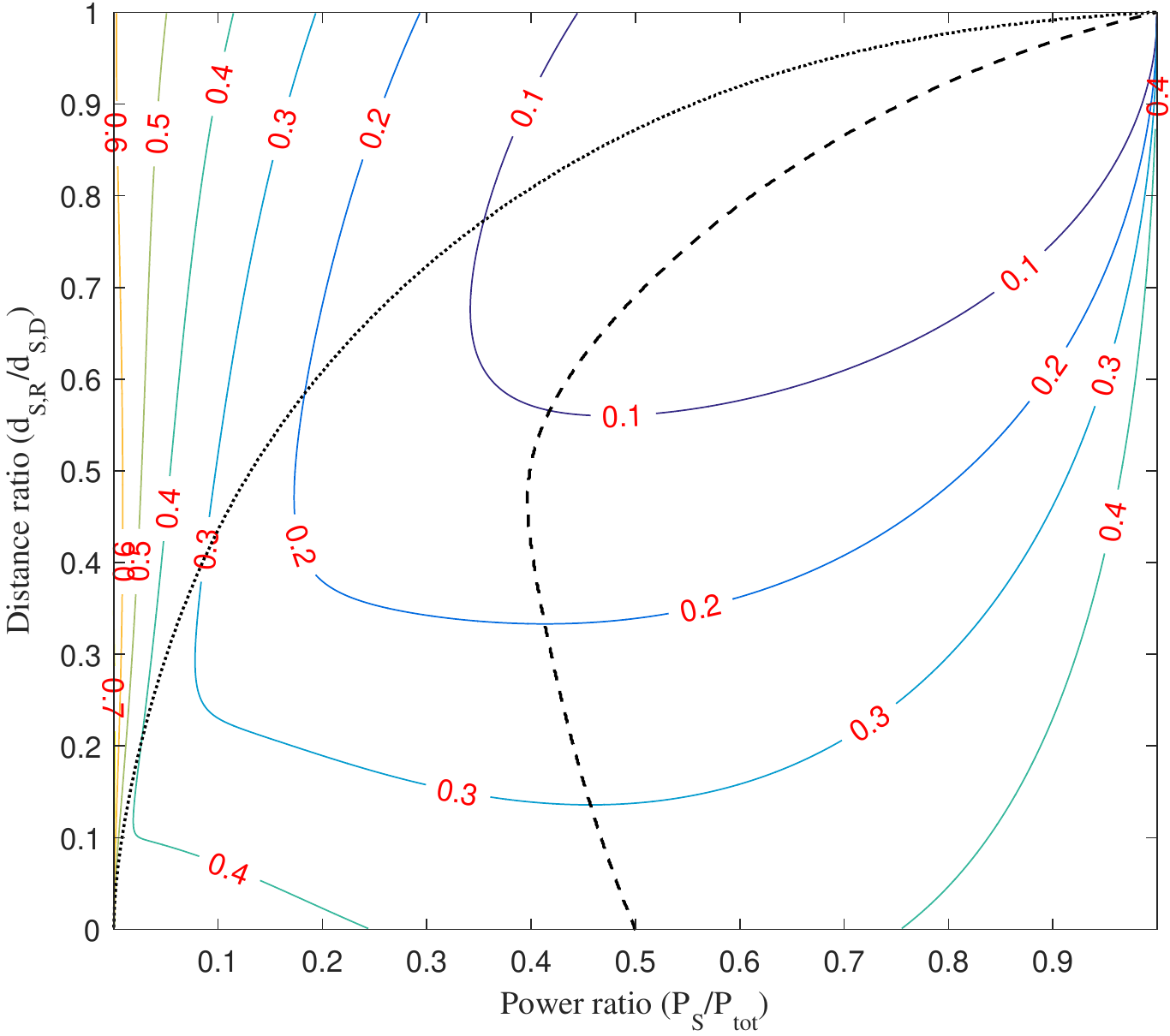,scale=0.59,angle=0}
\end{center}
\caption{The proposed system outage probability contour with $R=2$ bps/Hz, $\epsilon=1$, $\sigma^{2}_{\mathrm{R,R}}=0.1$ and $P_{\mathrm{tot}}=10$ watt. The dashed line represents the optimized power when relay location is fixed, and the dotted line represents the optimized relay location when power is fixed. The numbers on the contour curves denote the outage probability values.
}\label{F3}
\end{figure}

In the case where $X\neq Y$ (i.e. $\frac{P_{\mathrm{S}}}{d^{2}_{\mathrm{S,D}}}\neq\frac{P_{\mathrm{R}}}{d^{2}_{\mathrm{R,D}}}$), since there is no direct relation between transmission power and relay location, the optimal 
solution can be obtained by fixing one variable and optimizing the other. Specifically, by fixing the relay location and letting $P_{\mathrm{R}}=P_{\mathrm{tot}}-P_{\mathrm{S}}$, the optimal power allocation can be obtained by finding the root of equation $\frac{\partial\mathcal{P}^{\mathrm{FD}}_{\mathrm{out}}(P_{\mathrm{S}})}{\partial P_{\mathrm{S}}}=0$. Similarly, by fixing the power allocation and letting $d_{\mathrm{R,D}}=d_{\mathrm{S,D}}-d_{\mathrm{S,R}}$, the optimal relay location placement can be obtained by finding the root of equation $\frac{\partial\mathcal{P}^{\mathrm{FD}}_{\mathrm{out}}(d_{\mathrm{S,R}})}{\partial d_{\mathrm{S,R}}}=0$. Unlike the case where $X=Y$, here, $\mathcal{P}^{\mathrm{FD}}_{\mathrm{out}}(P_{\mathrm{S}})$ w.r.t. $P_{\mathrm{S}}$ (or $\mathcal{P}^{\mathrm{FD}}_{\mathrm{out}}(d_{\mathrm{S,R}})$ w.r.t. $d_{\mathrm{S,R}}$) is not a convex function. In this case, by employing the intermediate value theorem \cite{Rudin1976}, we can show that equation $\frac{\partial\mathcal{P}^{\mathrm{FD}}_{\mathrm{out}}(P_{\mathrm{S}})}{\partial P_{\mathrm{S}}}=0$ within the interval $[0,P_{\mathrm{tot}}]$ (or equation $\frac{\partial\mathcal{P}^{\mathrm{FD}}_{\mathrm{out}}(d_{\mathrm{S,R}})}{\partial d_{\mathrm{S,R}}}=0$ within the interval $[0,d_{\mathrm{S,D}}]$) has at least one root. Then, the bisection search method can be employed to find the local optimal point, where around fifteen iterations can lead to convergence. In Fig.~\ref{F3}, we show that the obtained local optimal point can offer better performances than the equal power allocation and the mid-distance relay location placement.    
 
Fig.~\ref{F3} depicts the outage probability contour for our proposed scheme with different transmission powers and relay locations setup. 
As shown in the figure, the simple equal power allocation or mid-distance relay location placement may not lead to the optimal outage probability. In addition, considering $P_{\mathrm{S}}/P_{\mathrm{tot}}=0.1$, a better outage probability can be obtained when the FD relay node gets closer to the source node. On the other hand, due to the total power constraint, larger $P_{\mathrm{S}}/P_{\mathrm{tot}}$ leads to smaller self-interference effect and higher decoding capability at the FD relay node. Thus, a better outage probability can be obtained by moving the FD relay node closer to the destination node to guarantee better R-D link quality.

{\em Remark 2:} In practical systems, the source and the FD relay nodes will have separate power constraints. In this case, by fixing the location of the FD relay node, the intuition for optimal power solution can be found as follow: Considering the outage expression in \eqref{eq21}, the outage probability is a monotonically decreasing function of $P_\mathrm{S}$ for any given $P_{\mathrm{R}}$. Thus, the optimal $P_{\mathrm{S}}$ is its maximum allowable power constraint. On the other hand, by fixing $P_{\mathrm{S}}$, the optimal $P_{\mathrm{R}}$ can be found by finding the root of equation $\frac{\partial\mathcal{P}^{\mathrm{FD}}_{\mathrm{out}}(P_{\mathrm{R}})}{\partial P_{\mathrm{R}}}=0$ with the help of the above discussed numerical search method. Then, the FD relay node needs to compare the optimal $P_{\mathrm{R}}$ with its maximum allowable power constraint, and select the smaller one as its actual transmitted power.   

%%%%%%%%%%%%%%%%%%%%%%%%%%%%%%%%%%%%%%%%%%%%%%
\subsection{Comparison with HD Relaying Protocol}
Due to self-interference effects, the symbol-level selective FD relaying may not always lead to a better performance in comparison with the one with HD relaying. In this case, it is worth formulating the corresponding outage expression of the symbol-level selective HD relaying, and identifying the condition that the symbol-level selective FD relaying can lead to a better performance.

In order to make a fair comparison, the outage derivation for the symbol-level selective HD relaying case should half the capacity (or double the target rate) by comparing with the ones for the FD relaying case. In addition, due to time-orthogonal transmissions, there is no self-interference effects in the HD relaying case. Thus, the probability that a symbol is selected at the HD relay node is equal to $\mathcal{P}_{0}$ as shown in Section IV-B. Then, due to the statistical identity and independence, the average probability of the selected symbols per frame $\mathcal{P}_{C}$ should be equal to $\mathcal{P}_{0}$. Based on the above description, the system outage probability of the symbol-level selective HD relaying can be derived as
\begin{equation}\label{eq21b}
\mathcal{P}^{\mathrm{HD}}_{\mathrm{out}}=\mathcal{P}_{0}\mathcal{P}^{\mathrm{HD}}_{\mathrm{FW}}+(1-\mathcal{P}_{0})\mathcal{P}^{\mathrm{HD}}_{\mathrm{Non-FW}},
\end{equation}
where
\begin{equation}\label{eq21c}
\mathcal{P}^{\mathrm{HD}}_{\mathrm{FW}}\triangleq\mathrm{Pr}\left[\frac{1}{2}C\left(P_{\mathrm{S}}|h_{\mathrm{S,D}}|^{2}+P_{\mathrm{R}}|h_{\mathrm{R,D}}|^{2}\right)< R\right],
\end{equation}
and
\begin{equation}\label{eq21d}
\mathcal{P}^{\mathrm{HD}}_{\mathrm{Non-FW}}\triangleq\mathrm{Pr}\left[\frac{1}{2}C\left(P_{\mathrm{S}}|h_{\mathrm{S,D}}|^{2}\right)< R\right].
\end{equation}
Unlike equation (19) in \cite{Laneman2004}, we didn't double the SNR of $\mathcal{P}^{\mathrm{HD}}_{\mathrm{Non-FW}}$ in \eqref{eq21d}. This is because, in this paper, we assume that there is no retransmission from the source node if the HD relay node fails its decoding process. Then, following the similar derivation process as \textit{Proposition 1}, the system outage probability of the symbol-level selective HD relaying can be expressed as
\begin{IEEEeqnarray}{ll}\label{eq21a}
\mathcal{P}^{\mathrm{HD}}_{\mathrm{out}}=\left\{
\begin{array}{l}
\mathcal{P}_{0}\left(1-\frac{e^{2R}-1+X}{X}e^{-\frac{e^{2R}-1}{X}}\right)+(1-\mathcal{P}_{0})\left(1-e^{-\frac{e^{2R}-1}{X}}\right),~~~~~~~~~~~~~~~~~~X=Y,\\
\mathcal{P}_{0}\bigg( 1-\frac{Y}{Y-X}e^{-\frac{e^{2R}-1}{Y}}+\frac{X}{Y-X}e^{-\frac{e^{2R}-1}{X}}\bigg)+(1-\mathcal{P}_{0})\left(1-e^{-\frac{e^{2R}-1}{X}}\right),~~~~X\neq Y.\\
\end{array}
\right.
\end{IEEEeqnarray}

Based on the derived outage expressions \eqref{eq21} and \eqref{eq21a}, the condition that the FD relaying case gives better outage performance than the HD relaying case is given by
\begin{equation}
\mathcal{P}^{\mathrm{FD}}_{\mathrm{out}}-\mathcal{P}^{\mathrm{HD}}_{\mathrm{out}}\geq0.
\end{equation}
In the next section, numerical results are provided to find the outage performance boundary between the FD relaying case and the HD relaying case with different transmission target rates and relay locations.

%%%%%%%%%%%%%%%%%%%%%%%%%%%%%%%%%%%%%%%%%%%%%%%%%%%%%%%%%%%%%%%%%%%%%%%%%%%%%%%%%%%%%%%%%%%%%%%%%%%%%%%%%%%%%%%%

%%%%% V. Simulation Results %%%%%%%

%%%%%%%%%%%%%%%%%%%%%%%%%%%%%%%%%%%%%%%%%%%%%%%%%%%%%%%%%%%%%%%%%%%%%%%%%%%%%%%%%%%%%%%%%%%%%%%%%%%%%%%%%%%%%%%%
\section{Numerical and Simulation Results}
%Para. 1 introduces the simulation environment as well as your experimental design.
In this section, we evaluate our proposed symbol-level selective FD relaying scheme in comparison with HD relaying case and two classic S-DF relaying protocols in terms of outage probability. In addition, we also provide the BER performances to further evaluate our proposed scheme. We assume all channel links are generated as independent block Rayleigh fading, which remain static over each time slot. $L=20$ frames are transmitted via $L+1$ time slots, and each frame conveys $M=512$ information bits. We define SNR as the transmission signal power to noise power ratio. The quadrature phase-shift keying (QPSK) modulation is used, so that the corresponding square deviation error $\varepsilon$ for the symbol-level selection is set to 0.5. Turbo-like channel coding is also considered for the BER performance analysis. The results are computed on average over 1000 independent channel realizations. Tab.~\ref{tab2} summarizes the parameters configuration for the discussed experiments.
\begin{table}[t]
\center
{\small
\caption{Summarizes Experiments settings in the paper}\label{tab2}
\begin{tabular}{ |p{1.8cm}|p{2.4cm}||p{1.8cm}|p{2.4cm}|}
\hline
\multicolumn{4}{|c|}{General Settings} \\
\hline
\textbf{Parameter} & \textbf{Value} & \textbf{Parameter} & \textbf{Value}\\
\hline
$L$ & 20 & $M$ & 512 \\
\hline
$d_{\mathrm{S,R}}$~$(L1)$ & 0.4$d$ & $d_{\mathrm{R,D}}$~$(L1)$ & 0.6$d$\\
\hline
$d_{\mathrm{S,R}}$~$(L2)$ & 0.8$d$ & $d_{\mathrm{R,D}}$~$(L2)$ & 0.2$d$\\
\hline
$\varepsilon$ & 0.5 &  & \\
\hline
\multicolumn{4}{|c|}{\textit{Experiment 1}} \\
\hline
\textbf{Parameter} & \textbf{Value} & \textbf{Parameter} & \textbf{Value}\\
\hline
$R$ & 1 \& 2  bps/Hz & $P_{\mathrm{S}}$ & 0.5$P_{\mathrm{tot}}$\\
\hline
$P_{\mathrm{R}}$ & 0.5$P_{\mathrm{tot}}$ & \\
\hline
\multicolumn{4}{|c|}{\textit{Experiment 2}} \\
\hline
\textbf{Parameter} & \textbf{Value} & \textbf{Parameter} & \textbf{Value}\\
\hline
$R$ & 2  bps/Hz & $\sigma^{2}_{\mathrm{R,R}}$ & 0.01 \& 1 \\
\hline
$\Gamma_{\mathrm{T}}$ & 3 (not in dB) & & \\
\hline
\multicolumn{4}{|c|}{\textit{Experiment 3}} \\
\hline
\textbf{Parameter} & \textbf{Value} & \textbf{Parameter} & \textbf{Value}\\
\hline
$\Gamma_{\mathrm{T}}$ & 3 (not in dB) & $\sigma^{2}_{\mathrm{R,R}}$ & 0 \& 0.01 \& 1 \\
\hline
$k$ & 0.5 & $G$ & $([3,2])_{8}$ \\
\hline
\end{tabular}}
\end{table}

\textit{Experiment 1:} This experiment aims to compare the FD relaying case with the HD relaying case in terms of outage probability. Both cases are implementing our proposed symbol-level selection method at the relay node. In addition, equal power allocation and two relay locations are considered. Specifically, given the distance of S-D link as $d_{\mathrm{S,D}}=d$, the two relay locations can be configured as: $L1$) $d_{\mathrm{S,R}}=0.4d$ and $d_{\mathrm{R,D}}=0.6d$; $L2$) $d_{\mathrm{S,R}}=0.8d$ and $d_{\mathrm{R,D}}=0.2d$. Then, the SNRs of the three channel links between three nodes can be approximated by $\mathrm{SNR}_{\mathrm{S,R}}=(\frac{d_{\mathrm{S,R}}}{d_{\mathrm{S,D}}})^{-v}\cdot\mathrm{SNR}_{\mathrm{S,D}}$ and $\mathrm{SNR}_{\mathrm{R,D}}=(\frac{d_{\mathrm{R,D}}}{d_{\mathrm{S,D}}})^{-v}\cdot\mathrm{SNR}_{\mathrm{S,D}}$ as in \cite{Anwar2012,Youssef2011}. In this section, we assume the path-loss exponent $v=2$ as in \cite{Liang2010}. Thus, the SNR relations in dB among different links can be approximated as: $L1$) $\mathrm{SNR}_{\mathrm{S,R}}=\mathrm{SNR}_{\mathrm{S,D}}+7.96$ and $\mathrm{SNR}_{\mathrm{R,D}}=\mathrm{SNR}_{\mathrm{S,D}}+4.44$; $L2$) $\mathrm{SNR}_{\mathrm{S,R}}=\mathrm{SNR}_{\mathrm{S,D}}+1.94$ and $\mathrm{SNR}_{\mathrm{R,D}}=\mathrm{SNR}_{\mathrm{S,D}}+13.98$. Here, $\mathrm{SNR}_{\mathrm{S,D}}$ is the half of total transmission power to the noise power ratio.

\begin{figure}[t] 
\begin{center}
\epsfig{figure=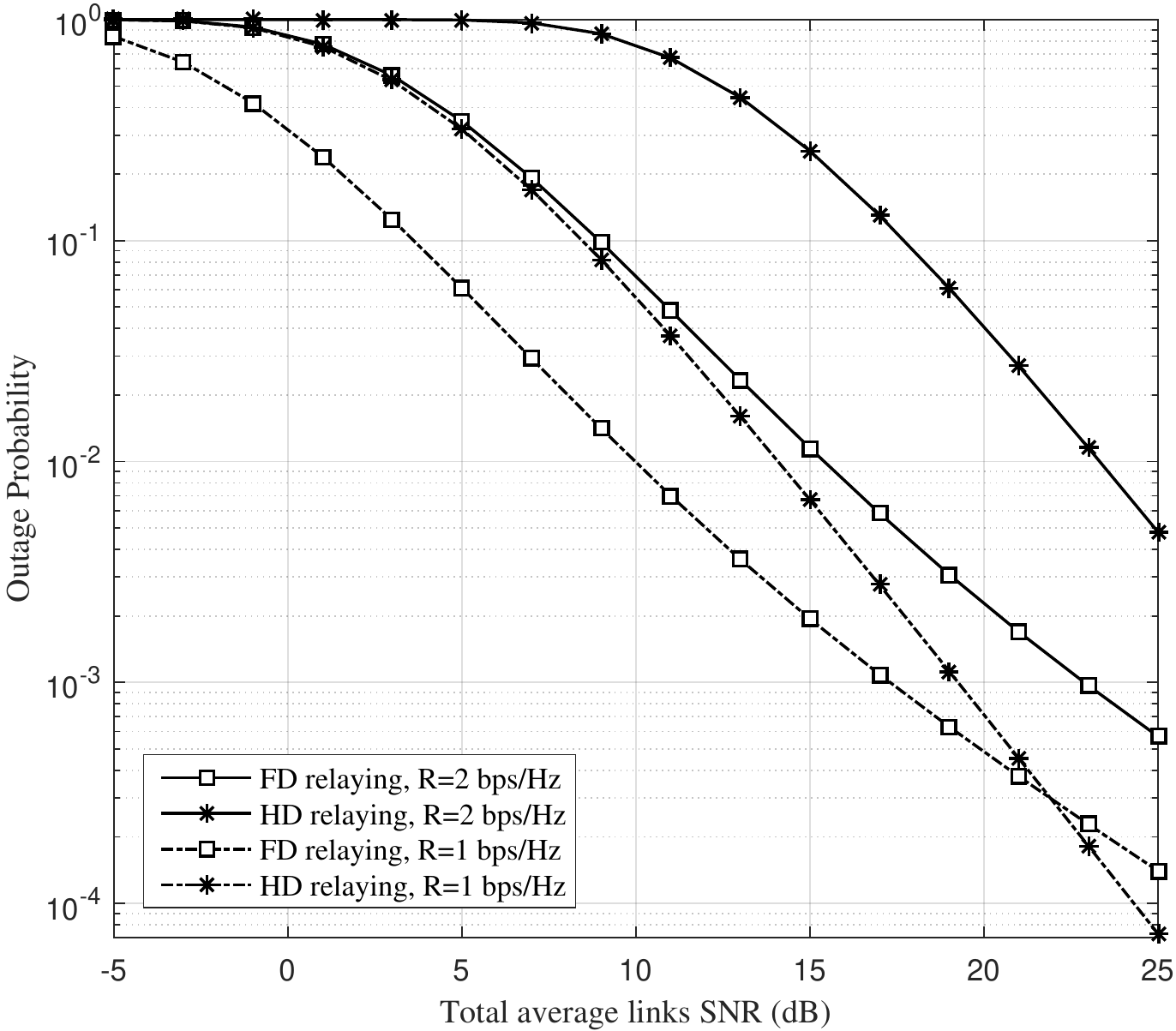,scale=0.59,angle=0}
\end{center}
\caption{Comparison between the FD relaying and the HD relaying in terms of outage probability with relay location $L1$, where the variance of self-interference channel at the FD relay node (i.e. $\sigma^{2}_{\mathrm{R,R}}$) is set to one, and the transmission target rate is set to $R=1$ bps/Hz and $R=2$ bps/Hz.}\label{F5}
\end{figure}

\begin{figure}[t] 
\begin{center}
\epsfig{figure=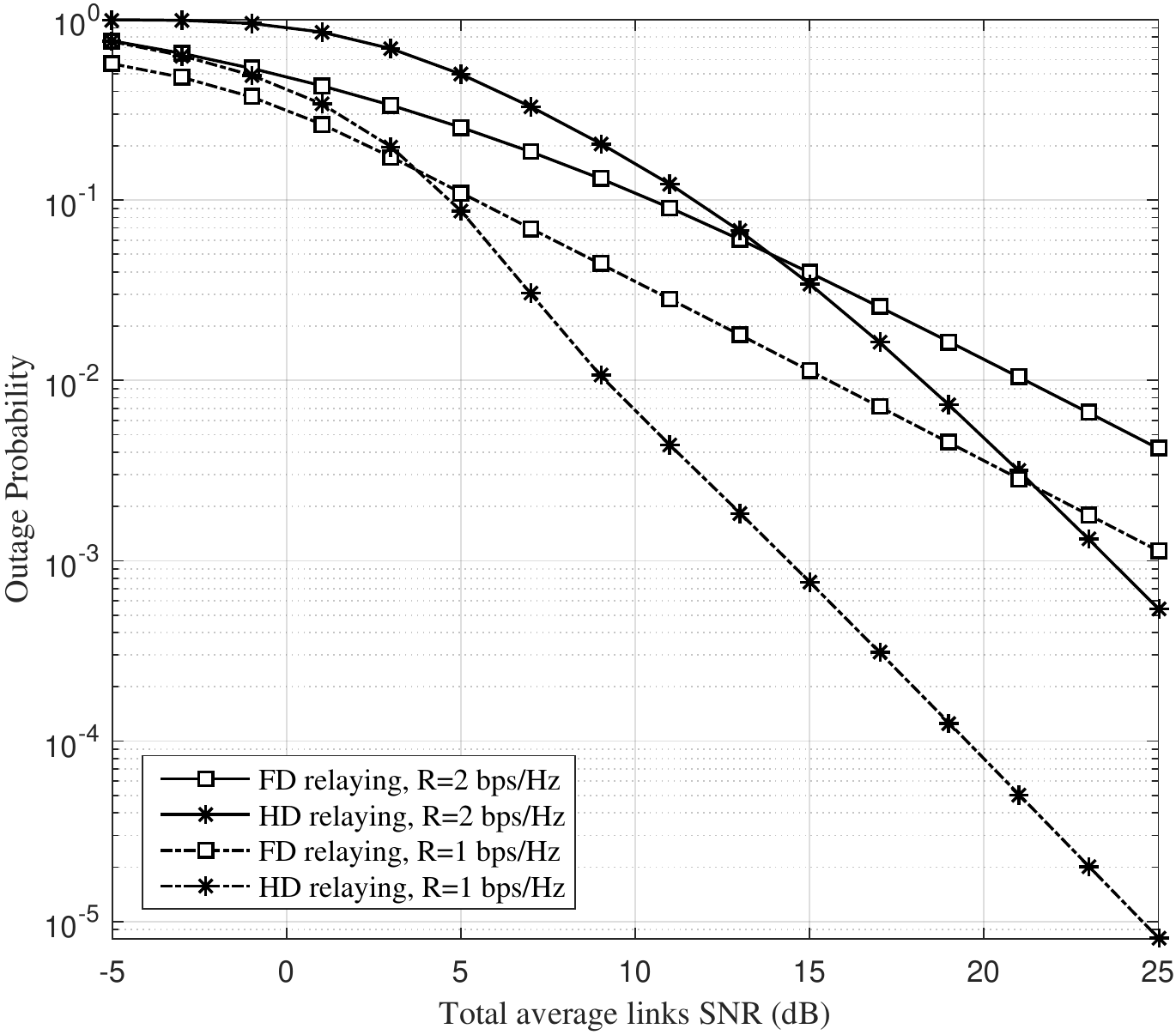,scale=0.59,angle=0}
\end{center}
\caption{Comparison between the FD relaying and the HD relaying in terms of outage probability with relay location $L2$, where the variance of self-interference channel at the FD relay node (i.e. $\sigma^{2}_{\mathrm{R,R}}$) is set to one, and the transmission target rate is set to $R=1$ bps/Hz and $R=2$ bps/Hz.}\label{F6}
\end{figure}

Fig.~\ref{F5} and Fig.~\ref{F6} provide the outage probability performances versus the total average links SNR with two relay locations, respectively. 
Here, the total average links SNR denotes the source node's SNR plus the relay node's SNR divided by two. As shown in Fig.~\ref{F5}, the FD relaying outperforms the HD relaying when the total average SNR is below 22 dB in the case $R=1$ bps/Hz. This is because with relay location $L1$ the decoding capability of the FD relay node can be guaranteed. In addition, the performance gap between the two relaying schemes is enlarged as the transmission target rate reaches to 2 bps/Hz. This means that the FD relaying can benefit the higher data rate transmission. As the relay node moves to the designation node, the performance trade-off between the FD relaying and the HD relaying should be considered. As shown in Fig.~\ref{F6}, for the target rate $R=1$ bps/Hz, the FD relaying scheme outperforms the HD relaying scheme only when the total average links SNR is less than 4dB. This boundary is increased to 14dB for the target rate $R=2$ bps/Hz. This is because self-interference becomes the dominant factor in the FD relaying case. 

\begin{figure}[t] 
\begin{center}
\epsfig{figure=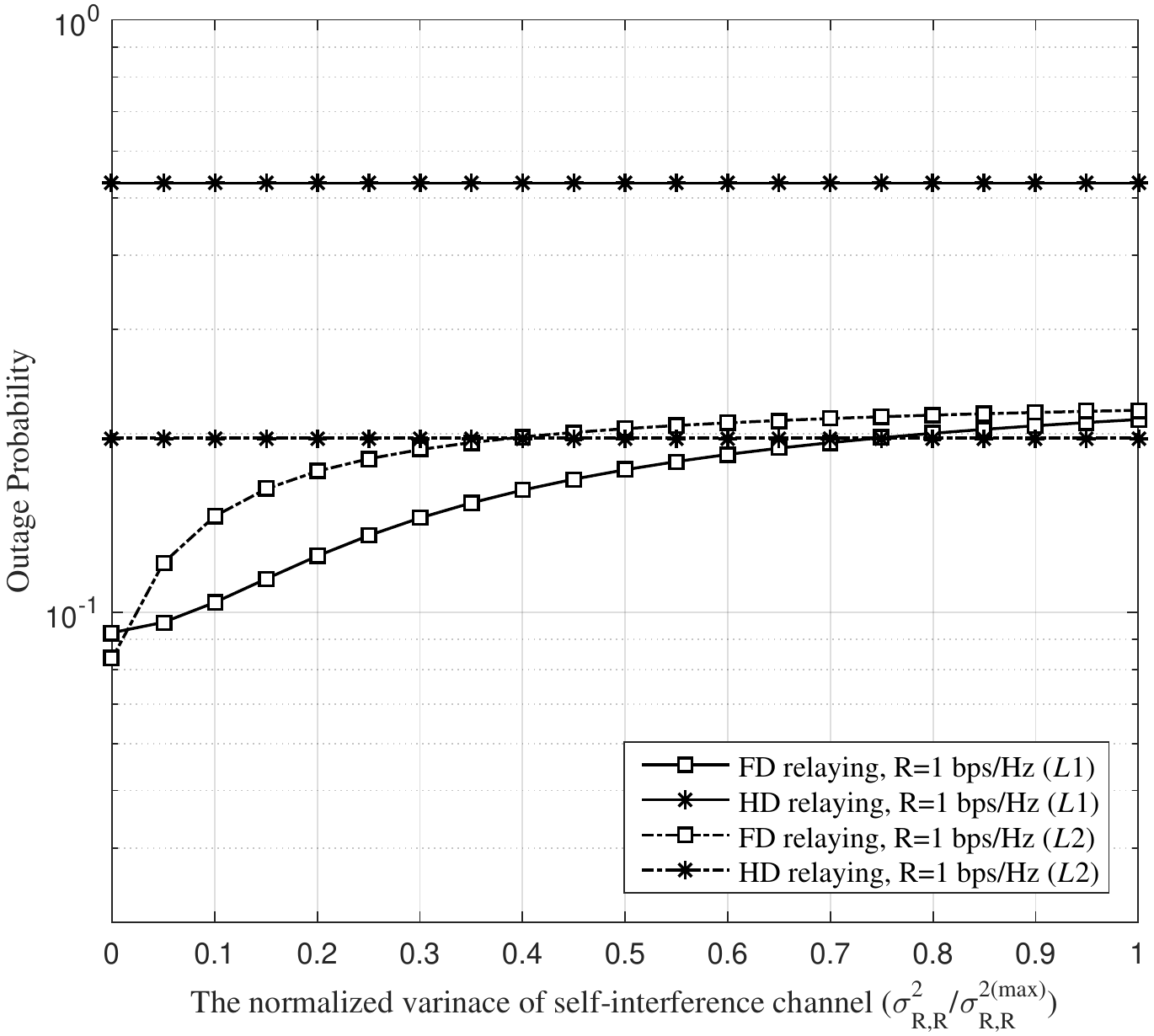,scale=0.59,angle=0}
\end{center}
\caption{Outage probability vs. the normalized variance of self-interference channel $\sigma^{2}_{\mathrm{R,R}}/\sigma^{2(\mathrm{max})}_{\mathrm{R,R}}$ with two relay locations, where the average links SNR is fixed to 3 dB, and the transmission target rate is set to $R=1$ bps/Hz.}\label{add01}
\end{figure}

\begin{figure}[t] 
\begin{center}
\epsfig{figure=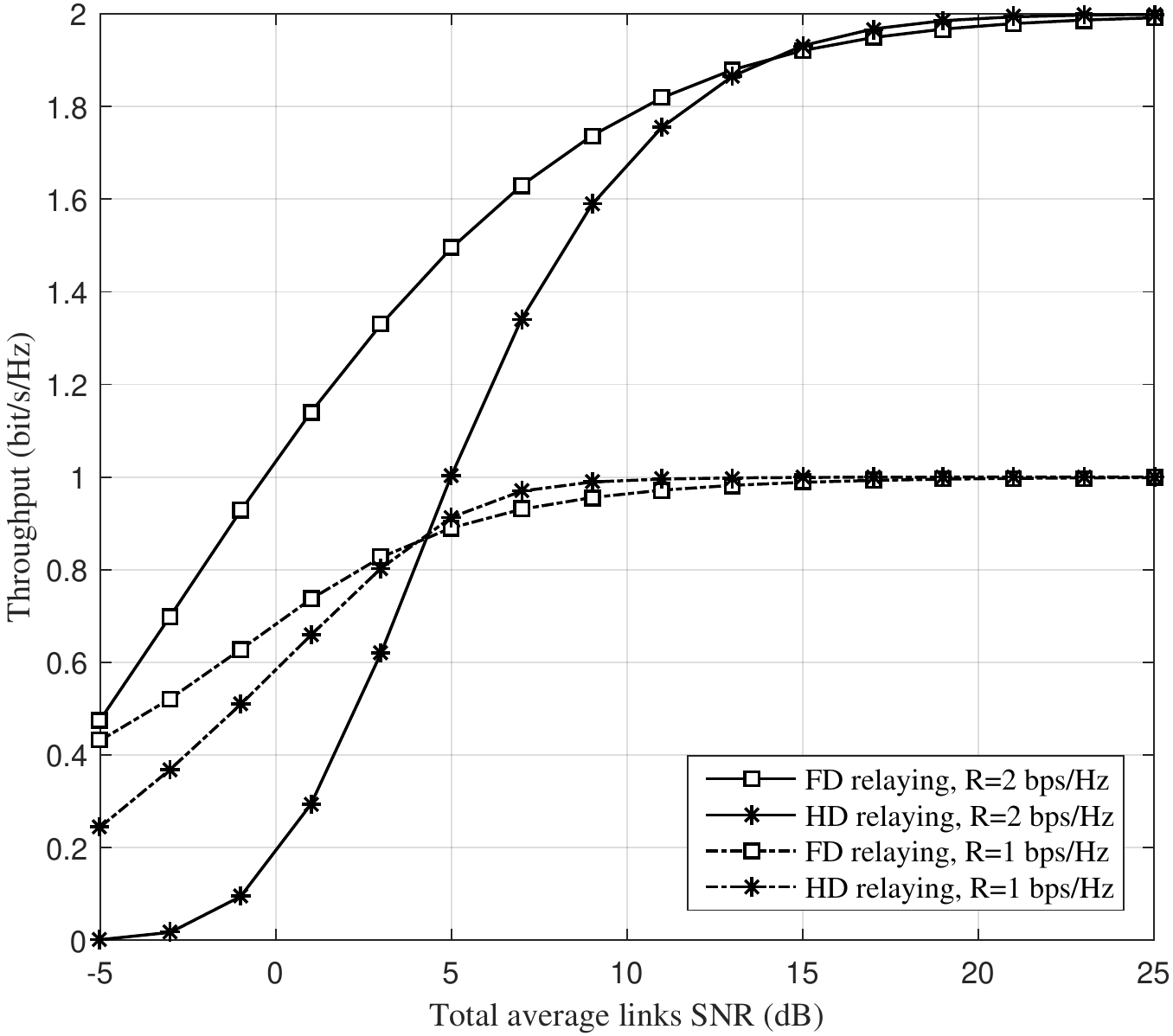,scale=0.59,angle=0}
\end{center}
\caption{Throughput vs. the total average links SNR with two transmission target rates, where the variance of self-interference channel at the FD relay node (i.e. $\sigma^{2}_{\mathrm{R,R}}$) is set to one, and relay location $L2$ is used as an example.}\label{add02}
\end{figure}

In this experiment, we also provide Fig.~\ref{add01} and Fig.~\ref{add02} to respectively demonstrate the outage performances versus different self-interference levels and the throughput performances versus different total average links SNR.
In Fig.~\ref{add01}, we assume the maximum variance of self-interference channel, i.e., $\sigma^{2(\mathrm{max})}_{\mathrm{R,R}}$, is equal to five. The outage performance for the FD relaying case is decreasing accompanied by increasing the normalized variance of self-interference channel $\sigma^{2}_{\mathrm{R,R}}/\sigma^{2(\mathrm{max})}_{\mathrm{R,R}}$. Here, for relay location $L1$, the FD relaying case outperforms the HD relaying case for entire normalized variances of self-interference. This is because the decoding capability of the FD relay node can be guaranteed in this case. For relay location $L2$, the FD relaying case outperforms the HD relaying case when $\sigma^{2}_{\mathrm{R,R}}/\sigma^{2(\mathrm{max})}_{\mathrm{R,R}}$ is below 0.4. This is because the self-interference dominates the decoding capability for relay location $L2$ when $\sigma^{2}_{\mathrm{R,R}}/\sigma^{2(\mathrm{max})}_{\mathrm{R,R}}$ is above 0.4. Then, as shown in Fig.~\ref{add02}, the throughput performances for both FD relaying and HD relaying are increasing accompanied by increasing the total average links SNR, and their performances convergence at high SNR range. This is because the target rate can be guaranteed for both cases with high transmission power. In addition, for the low SNR range, as we expected, the FD relaying outperforms the HD relaying for both target rates.

%%%%%%%%%%%%%%%%%%%%%%%%%%%
\textit{Experiment 2:} In this experiment, assuming the relay node has FD capability, we compare our proposed symbol-level selection method with two classic S-DF relaying protocols in terms of outage probability. The above mentioned two relay location placements are also considered in this experiment. The two classic relaying protocols are: 1) \textit{CRC based S-DF}, where the FD relay node only helps if the received frame passes the CRC check; 2) \textit{Threshold based S-DF}, where the FD relay node only helps if its received SINR is larger than a pre-determined threshold, e.g., $\Gamma_{\mathrm{T}}=3$ in this experiment. In order to obtain the system outage probability expression of \textit{CRC based S-DF} based protocol, we need to replace $\mathcal{P}_{\mathrm{S}}$ in \eqref{eq17} with 
\begin{eqnarray}\label{eq17a}
\mathcal{P}^{\mathrm{CRC}}_{\mathrm{S}}&=&1-\mathrm{Pr}\left[C\left(\frac{P_{\mathrm{S}}|h_{\mathrm{S,R}}|^{2}}{P_{\mathrm{R}}|h_{\mathrm{R,R}}|^{2}+1}\right)<R\right]\nonumber\\
&=&\frac{1}{1+\frac{P_{\mathrm{R}}\sigma^{2}_{h_{\mathrm{R,R}}}(e^{R}-1)}{P_{\mathrm{S}}\sigma^{2}_{\mathrm{S,R}}}}\mathrm{e}^{-\frac{e^{R}-1}{P_{\mathrm{S}}\sigma^{2}_{h_{\mathrm{S,R}}}}},
\end{eqnarray}
and then follow the same steps as \eqref{eq18}, \eqref{eq19} and \eqref{eq20} in Section IV-A to formulate the system outage probability for \textit{CRC based S-DF} protocol. Here, \eqref{eq17a} is derived with the help of solving the outage probability on the condition that the S-R link channel gain is given. More detailed analysis refers to equation (4) and equation (5) in \cite{Kwon2010}. Similarly, we can formulate the system outage probability of \textit{Threshold based S-DF} based protocol by replacing $(e^{R}-1)$ with $\Gamma_{\mathrm{T}}$ in \eqref{eq17a}, and then following the same steps as \textit{CRC based S-DF} case.

\begin{figure}[t] 
\begin{center}
\epsfig{figure=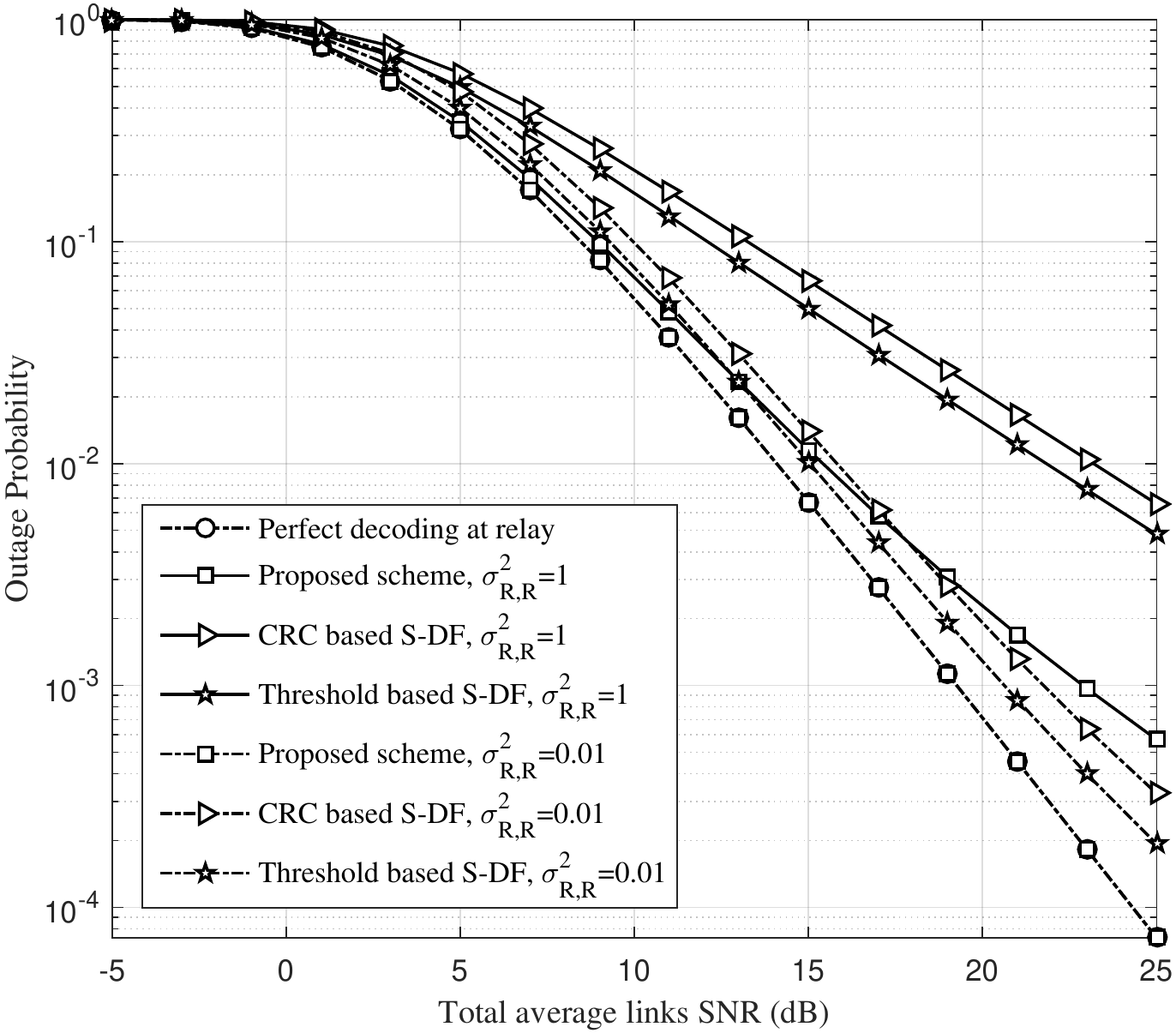,scale=0.59,angle=0}
\end{center}
\caption{Outage probability versus the total average links SNR for different relaying protocols with relay location $L1$, where the variance of self-interference channel at the FD relay node (i.e. $\sigma^2_{\mathrm{R,R}}$) is set to 1 and 0.01, respectively. Transmission target rate is set to $R=2$ bps/Hz, and the pre-determined SINR threshold is set to $\Gamma_{\mathrm{T}}=3$.}\label{F7}
\end{figure}

\begin{figure}[t] 
\begin{center}
\epsfig{figure=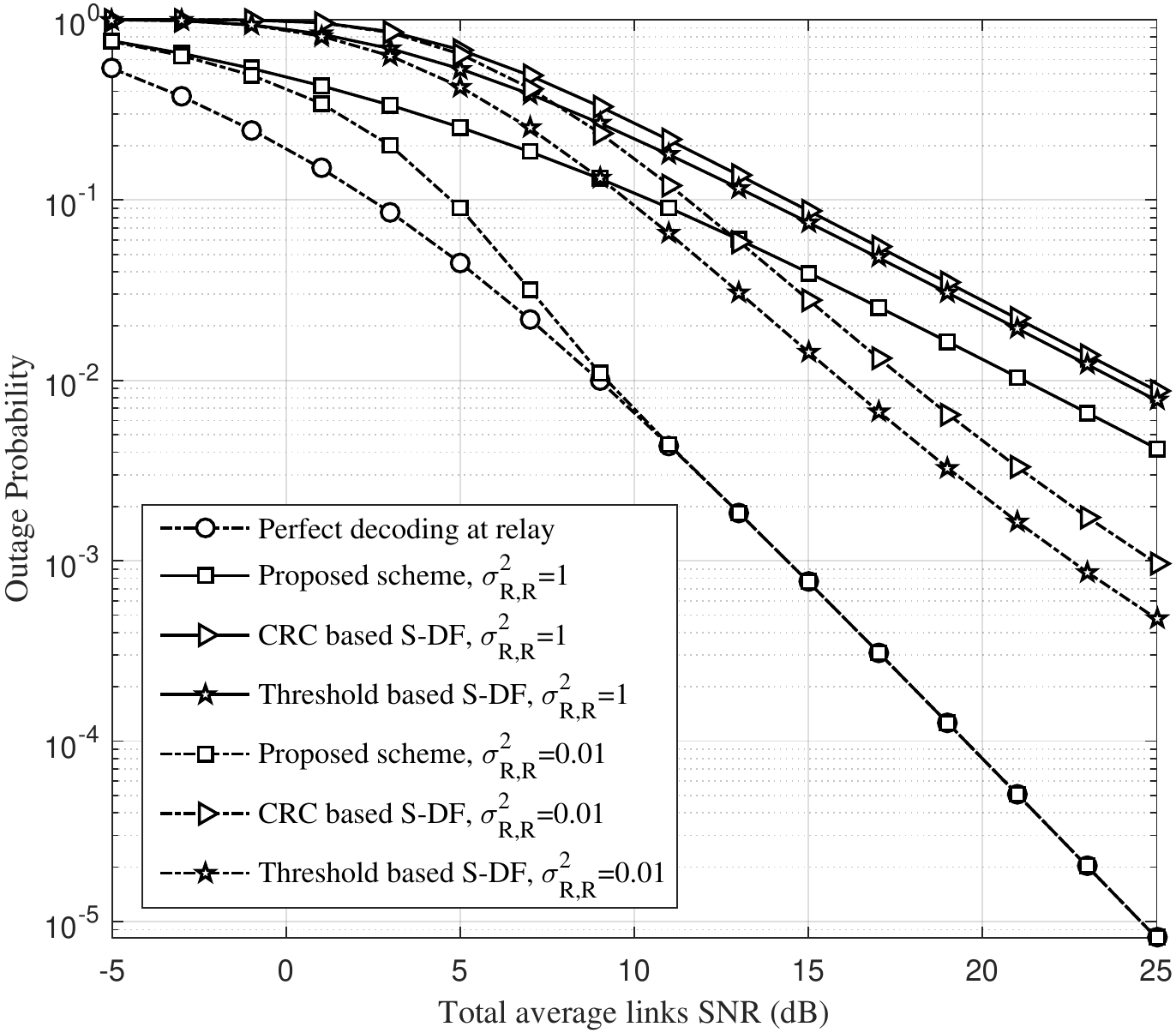,scale=0.59,angle=0}
\end{center}
\caption{Outage probability versus the total average links SNR for different relaying protocols with relay location $L2$, where the variance of self-interference channel at the FD relay node (i.e. $\sigma^2_{\mathrm{R,R}}$) is set to 1 and 0.01, respectively. Transmission target rate is set to $R=2$ bps/Hz, and the pre-determined SINR threshold is set to $\Gamma_{\mathrm{T}}=3$.}\label{F8}
\end{figure}

Fig.~\ref{F7} and Fig.~\ref{F8} give the outage probability performances versus the total average links SNR for different relaying schemes and self-interference levels.
As shown in the figures with two different relay locations, our proposed scheme outperforms both \textit{Threshold based S-DF} scheme and \textit{CRC based S-DF} scheme in two different self-interference levels (i.e. $\sigma^{2}_{\mathrm{R,R}}=1$ and $\sigma^{2}_{\mathrm{R,R}}=0.01$). Moreover, \textit{Threshold based S-DF} scheme has better outage performances than \textit{CRC based S-DF} scheme. This is because, for \textit{CRC based S-DF} scheme, only one or several erroneous bits would trigger a CRC failure and stop a significant number of correct bits to be forwarded to the destination node, resulting in diversity gain loss. In addition, we also provide \textit{Perfect decoding at relay} scheme, where the FD relay node can always perfectly decode its received signals. Comparing Fig.~\ref{F7} and Fig.~\ref{F8}, the outage performance of our proposed scheme in Fig.~\ref{F7} is almost overlap with the ones of \textit{Perfect decoding at relay} scheme. This is because the FD relay node with relay location $L1$ is close to the source node and high probability of successfully decoding at the FD relay node can be guaranteed. On the other hand, due to relay location $L2$ and self-interference effects, there is a performance gap between our proposed scheme and \textit{Perfect decoding at relay} scheme around low SNR range in Fig.~\ref{F8}.

\begin{figure}[t] 
\begin{center}
\epsfig{figure=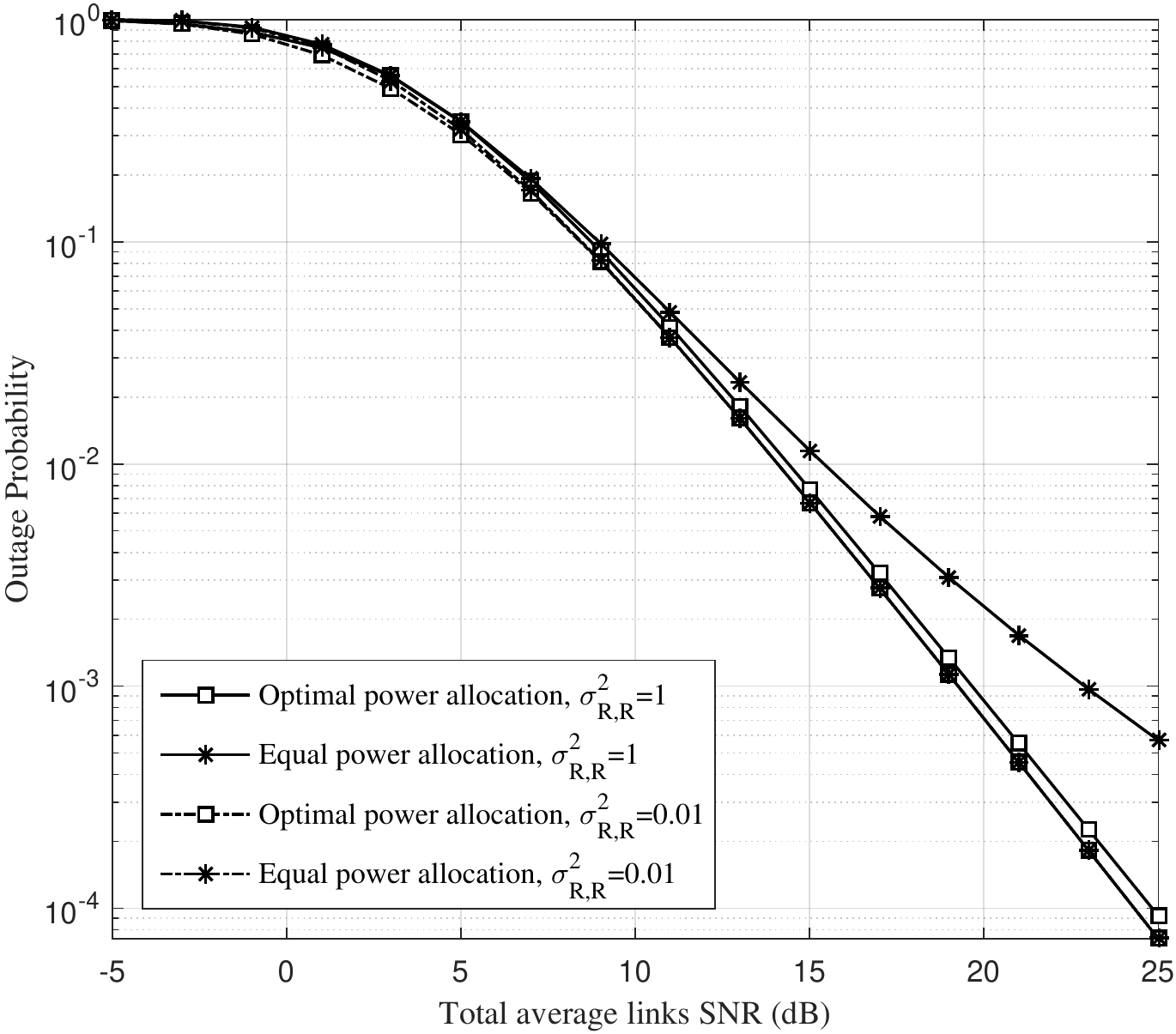,scale=0.59,angle=0}
\end{center}
\caption{Comparison between optimal power allocation and equal power allocation of our proposed scheme in terms of outage probability with relay location $L1$, where the variance of self-interference channel at the FD relay node (i.e. $\sigma^{2}_{\mathrm{R,R}}$) is set to 1 and 0.01, and the transmission target rate is set to $R=2$ bps/Hz.}\label{F9}
\end{figure}

\begin{figure}[t] 
\begin{center}
\epsfig{figure=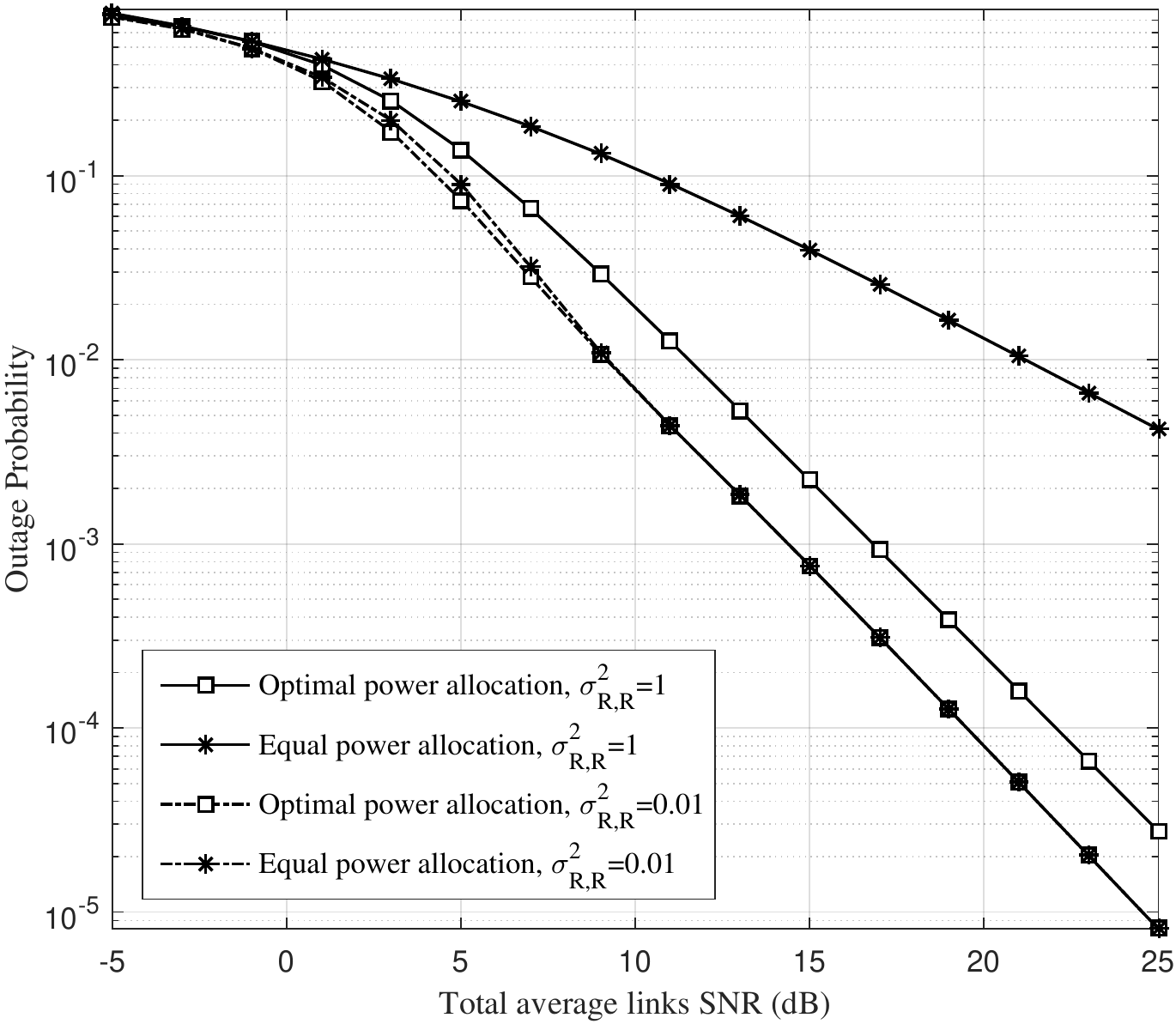,scale=0.59,angle=0}
\end{center}
\caption{Comparison between optimal power allocation and equal power allocation of our proposed scheme in terms of outage probability with relay location $L2$, where the variance of self-interference channel at the FD relay node (i.e. $\sigma^{2}_{\mathrm{R,R}}$) is set to 1 and 0.01, and the transmission target rate is set to $R=2$ bps/Hz.}\label{F10}
\end{figure}

Apart from the comparison among different relaying protocols, in this experiment, we also compare the optimal power allocation with the equal power allocation of our proposed scheme for the two relay locations in Fig.~\ref{F9} and Fig.~\ref{F10}, respectively.
As shown in Fig.~\ref{F9}, the outage probability with optimal power allocation outperforms the one with equal power allocation when the total average link SNR is above 7 dB in the case $\sigma^{2}_{\mathrm{R,R}}=1$. In the case $\sigma^{2}_{\mathrm{R,R}}=0.01$, equal power allocation can give nearly optimal performance due to the reduced self-interference effects. In comparison, Fig.~\ref{F10} provides a big performance gap between optimal power allocation and equal power allocation in the case $\sigma^{2}_{\mathrm{R,R}}=1$. This is because, with relay location $L2$, the decoding capability of the FD relay node is decreasing especially in the presence of strong self-interference. In this case, optimal power allocation can be exploited to provide a better performance. On the other hand, the performance gap is reducing accompanied by reducing the self-interference effects, e.g., $\sigma^{2}_{\mathrm{R,R}}=0.01$.

\textit{Experiment 3:} The objective of this experiment is to examine our proposed symbol-level selective method in the presence of a specific channel coding method. In this case, BER is used to evaluate the system performances. Specifically, a 1/2-rate serial concatenated convolutional code is used at both the source node and the FD relay node, where the first encoder is the non-recursive non-systematic convolutional code with a generator polynomial $G=([3,2])_{8}$, and the second encoder is the doped-accumulator with a doping rate equalling two. The destination node for all presented schemes incorporates the same modified MAP receiver as our proposed scheme. In addition, the relay location $L1$ is used as an example.

\begin{figure}[t] 
\begin{center}
\epsfig{figure=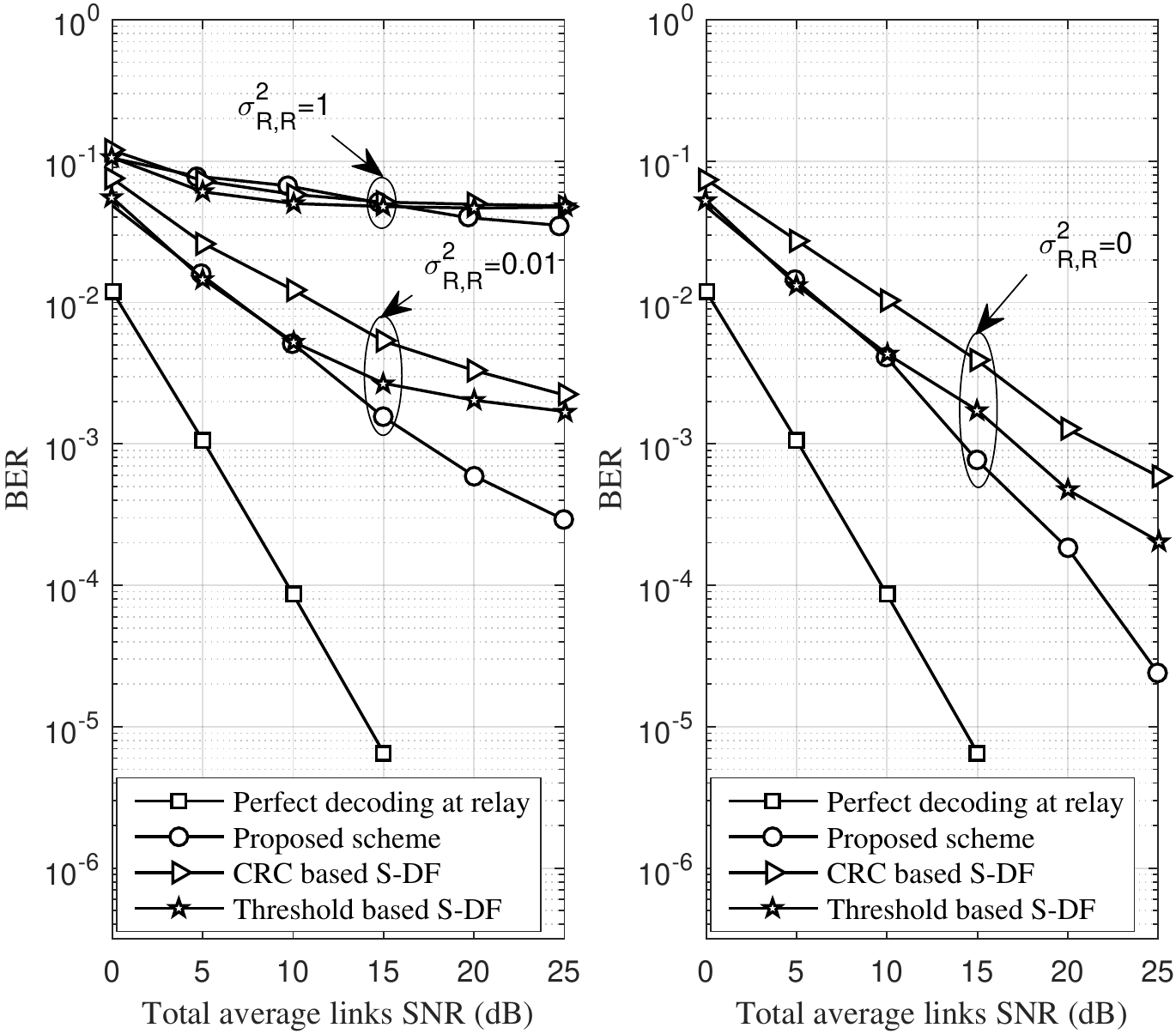,scale=0.59,angle=0}
\end{center}
\caption{BER versus the total average links SNR for different relaying protocols, where the variance of self-interference channel at the FD relay node (i.e. $\sigma^2_{\mathrm{R,R}}$) is set to 1, 0.01 and 0, respectively, and the pre-determined SINR threshold is set to $\Gamma_{\mathrm{T}}=3$.}\label{F11}
\end{figure}

Fig.~\ref{F11} gives BER performance versus the total average links SNR for different relaying protocols  with different self-interference levels. 
As shown in the figure, our proposed scheme provides better BER performances in comparison with \textit{Threshold based S-DF} scheme and \textit{CRC based S-DF} scheme especially for high SNR range. This BER performance trend is in line with the outage performances. Specifically, in the case $\sigma^2_{\mathrm{R,R}}=1$, our proposed scheme outperforms both \textit{Threshold based S-DF} scheme and \textit{CRC based S-DF} scheme when the total average links SNR is larger than 16 dB. In the case $\sigma^2_{\mathrm{R,R}}=0.01$, our proposed scheme and \textit{Threshold based S-DF} scheme outperform \textit{CRC based S-DF} scheme for the entire presented SNR range. On the other hand, our proposed scheme outperforms \textit{Threshold based S-DF} scheme when the SNR is larger than 10 dB. In the case $\sigma^2_{\mathrm{R,R}}=0$, the same performance trend is seen as the case $\sigma^2_{\mathrm{R,R}}=0.01$. In addition, unlike the outage performances, there is a large BER performance gap between our proposed scheme and \textit{Perfect decoding at relay} scheme. This is because optimal decoding method at the destination node was assumed when we derived the outage probabilities in Section IV-A. In contrast, for the BER performances, the practical low-complex turbo-like decoding method and the modified MAP detector degrade the system performances. 

\begin{figure}[t] 
\begin{center}
\epsfig{figure=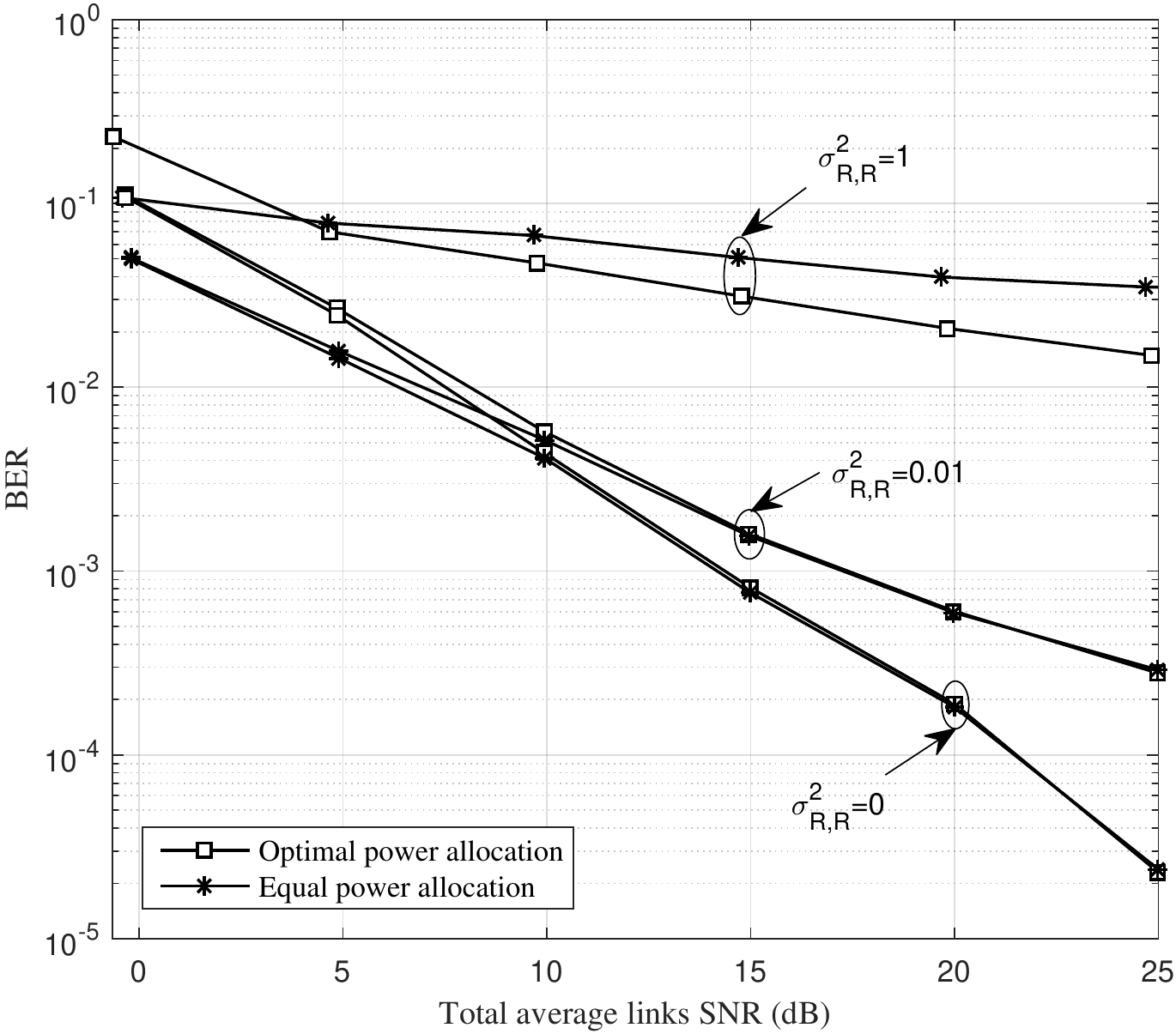,scale=0.59,angle=0}
\end{center}
\caption{Comparison between optimal power allocation and equal power allocation of our proposed scheme in terms of BER performances, where the variance of self-interference channel at the FD relay node (i.e. $\sigma^{2}_{\mathrm{R,R}}$) is set to 1, 0.01 and 0, respectively.}\label{F12}
\end{figure}

Fig.~\ref{F12} gives the comparison between optimal power allocation and equal power allocation of our proposed scheme in terms of BER performances.
As shown in the figure, the BER performances with optimal power allocation outperform the ones with equal power allocation when the self-interference at the FD relay node is set to $\sigma^{2}_{\mathrm{R,R}}=1$. This is because, in this case, the source node requires more transmission power to guarantee the quality of decoding process at the FD relay node. For the other self-interference cases, equal power allocation is nearly equivalent to optimal power allocation. It is worth noting that, for low average SNR range, we observe that the BER performance with optimal power allocation is worse than the ones with equal power allocation. This is because that the imperfect detection and decoding may affect the BER performances, in addition, the optimization is in terms of the outage probabilities.

%%%%%%%%%%%%%%%%%%%%%%%%%%%%%%%%%%%%%%%%%%%%%%%%%%%%%%%%%%%%%%%%%%%%%%%%%%%%%%%%%%%%%%%%%%%%%%%%%%%%%%%%%%%%%%%%%%%%%%%%%%%%%%%%%%%%%%%%%%%%%%%%%%%%%%%%

                                   %%%VI. Conclusion%%%

%%%%%%%%%%%%%%%%%%%%%%%%%%%%%%%%%%%%%%%%%%%%%%%%%%%%%%%%%%%%%%%%%%%%%%%%%%%%%%%%%%%%%%%%%%%%%%%%%%%%%%%%%%%%%%%%%%%%%%%%%%%%%%%%%%%%%%%%%%%%%%%%%%%%%%%%
\section{Conclusion}
In this paper, we proposed a symbol-level selective transmission for FD relaying network. The proposed scheme predicts the correctly decoded symbols at the FD relay node using the square deviation method. Then, the destination node implements the modified MAP detection algorithm to cancel the inter-frame interference and identify the positions of discarded symbols at the FD relay node. Furthermore, the outage probability of the proposed scheme has been derived and compared with HD relaying case. In addition, the power allocation and relay location optimizations have also been analysed based on the derived outage probabilities. The results have shown that, our proposed scheme outperforms the classic CRC based S-DF relaying and threshold based S-DF relaying schemes in terms of both outage and BER. In addition, our proposed scheme with optimal power allocation outperforms the scheme with equal power allocation, especially when the self-interference at the FD relay node is strong and the FD relay node is close to the destination node. For the case where the self-interference at the FD relay node is weak, equal power allocation leads to near optimal performances for different relay locations.   

%%%%%%%%%%%%%%%%%%%%%%%%%%%%%%%%%%%%%%%%%%%%%%%%%%%%%%%%%%%%%%%%%%%%%%

                         %%%Acknowledgment%%%

%%%%%%%%%%%%%%%%%%%%%%%%%%%%%%%%%%%%%%%%%%%%%%%%%%%%%%%%%%%%%%%%%%%%%%
%\section*{Acknowledgment}
%The authors would like to thank...% the Editor Prof. and anonymous reviewers for their efficient review process and constructive comments.
%%%%%%%%%%%%%%%%%%%%%%%%%%%%%%%%%%%%%%%%%%%%%%%%%%%%%%%%%%%%%%%%%%%%%%
\section*{Appendix A\\Accuracy Analysis of The Symbol-Level Selection Method}
As shown in \eqref{eq11} and \eqref{eq13}, a symbol, e.g., $\hat{\mathbf{x}}^{(m)}_{\mathrm{R}}(l)$, is assumed to be decoded and selected correctly if $\mathbf{W}_{\mathrm{R}}(l)\tilde{\mathbf{y}}^{(m)}_{\mathrm{R}}(l)$ is closer to $\hat{\mathbf{x}}^{(m)}_{\mathrm{R}}(l)$, i.e., $\Delta_m(l)\leq\varepsilon$. However, there is possibility of that $\hat{\mathbf{x}}^{(m)}_{\mathrm{R}}(l)$ has been decoded incorrectly and $\mathbf{W}_{\mathrm{R}}(l)\tilde{\mathbf{y}}^{(m)}_{\mathrm{R}}(l)$ is still closer to $\hat{\mathbf{x}}^{(m)}_{\mathrm{R}}(l)$. In this case, the symbol will be selected incorrectly. In this appendix, we prove that such event can rarely happen.%, especially for high modulation schemes.

In detail, following $Q$-ary modulation, we define $\mathcal{P}_{i}$ as the probability of that $\hat{\mathbf{x}}^{(m)}_{\mathrm{R}}(l)$ is matching the $i^{\mathrm{th}}$ constellation point, where $\sum^{Q}_{i=1}\mathcal{P}_{i}=1$. In addition, we also define $\mathcal{P}'_{i}$ as the probability of that $\mathbf{W}_{\mathrm{R}}(l)\tilde{\mathbf{y}}^{(m)}_{\mathrm{R}}(l)$ is closer to the $i^{\mathrm{th}}$ constellation point, where $\sum^{Q}_{i=1}\mathcal{P}'_{i}=1$. In this case, if the $i^{\mathrm{th}}$ constellation point represents the correctly decoded symbol, the probability of that a symbol has been selected correctly at the FD relay node can be formulated as
\begin{equation}\label{eqB1}
\mathcal{P}_{\mathrm{R}}=\mathcal{P}_{i}\mathcal{P}'_{i},~i\in[1,\ldots,Q].
\end{equation}
On the other hand, the probability of that a symbol has been selected incorrectly can be formulated as
\begin{equation}\label{eqB2}
\mathcal{P}_{\mathrm{W}}=\sum_{j\neq i}\mathcal{P}_{j}\mathcal{P}'_{j},~i,j \in[1,\ldots,Q].
\end{equation}
As we increase the modulation order $Q$, it is possible to keep $\mathcal{P}_{i}$ and $\mathcal{P}'_{i}$ constant (or much larger than the other probabilities) by increasing the transmit power on the source node. Under this circumstances, because $\sum^{Q}_{i=1}\mathcal{P}_{i}=1$ and $\sum^{Q}_{i=1}\mathcal{P}'_{i}=1$, it is easy to prove that
\begin{equation}\label{eqB3}
\lim_{Q\rightarrow\infty}\mathrm{Pr}[\mathcal{P}_{i}\mathcal{P}'_{i}-\sum_{j\neq i}\mathcal{P}_{j}\mathcal{P}'_{j}\gg0]=1,~i,j\in\{1,\ldots,Q\}.
\end{equation}
Thus, the accuracy of the symbol-level selection method is guaranteed, especially for high modulation schemes and high transmit power on the source node. It is worth noting that error floor of the channel decoding process may affect the accuracy of our proposed symbol-level selection method.

%%%%%%%%%%%%%%%%%%%%%%%%%%%%%%%%%%%%%%%%%%%%%%%%%%%%%%%%%%%%%%%%%%%%%%
\section*{Appendix B\\Derivation of (26)}
We start with the case where the FD relay node can decode its received symbols correctly. In this case, based on \eqref{eq04}, the system outage probability can be approximated as 
\begin{equation}\label{eqA1}
\mathcal{P}^{\mathrm{FD}}_{\mathrm{FW}}\approx\mathrm{Pr}\left[C\left(P_{\mathrm{S}}|h_{\mathrm{S,D}}|^{2}+P_{\mathrm{R}}|h_{\mathrm{R,D}}|^{2}\right)< R\right],
\end{equation}
where $C(x)\triangleq\log(1+x)$ denotes Shannon rate, and $R$ is the transmission target rate. For the sake of simplicity, the natural logarithm is used to derive the outage probabilities, however, the actual Shannon rate should be based on binary logarithm. The similar expression as \eqref{eqA1} can also be found in \cite{Khafagy2015}, where the channel there, as we mentioned in Section II, was assumed to change each super-block but not each frame, which is different from our case. By considering Rayleigh fading for both S-D and R-D links, $P_{\mathrm{S}}|h_{\mathrm{S,D}}|^{2}$ and $P_{\mathrm{R}}|h_{\mathrm{R,D}}|^{2}$ follow exponential distribution with \textit{rate parameter} $1/(P_{\mathrm{S}}\sigma^{2}_{h_{\mathrm{S,D}}})$ and $1/(P_{\mathrm{R}}\sigma^{2}_{h_{\mathrm{R,D}}})$, respectively. Let $X\triangleq P_{\mathrm{S}}\sigma^{2}_{h_{\mathrm{S,D}}}$, $Y\triangleq P_{\mathrm{R}}\sigma^{2}_{h_{\mathrm{R,D}}}$, and $f_{\mathrm{X}}$, $f_{\mathrm{Y}}$ be their respective densities. Then, the p.d.f. of $Z=X+Y$ can be expressed as
\begin{eqnarray}\label{eqA2}
f_{\mathrm{Z}}(z)&=&\int^{+\infty}_{-\infty}f_{\mathrm{X}}(z-y)f_{\mathrm{Y}}(y)dy\nonumber\\
&=&\frac{1}{XY}e^{-\frac{z}{X}}\int^{z}_{0}e^{\frac{Y-X}{XY}y}dy
\end{eqnarray}
Here, if $X=Y$,
\begin{equation}\label{eqA3}
f_{\mathrm{Z}}(z)=\frac{z}{X^{2}}e^{-\frac{z}{X}},
\end{equation}
and if $X\neq Y$,
\begin{equation}\label{eqA4}
f_{\mathrm{Z}}(z)=\frac{1}{Y-X}\left(e^{-\frac{z}{Y}}-e^{-\frac{z}{X}}\right).
\end{equation}
With the distribution expressions in \eqref{eqA3} and \eqref{eqA4}, the system outage can be calculated as, if $X=Y$,
\begin{equation}\label{eqA5}
\mathcal{P}^{\mathrm{FD}}_{\mathrm{FW}}=1-\left(\frac{e^{R}-1+X}{X}\right)e^{-\frac{e^{R}-1}{X}},
\end{equation}
and if $X\neq Y$,
\begin{equation}\label{eqA6}
\mathcal{P}^{\mathrm{FD}}_{\mathrm{FW}}=1-\left(\frac{Ye^{-\frac{e^{R}-1}{Y}}-Xe^{-\frac{e^{R}-1}{X}}}{Y-X}\right).
\end{equation}

On the other hand, if the FD relay node doesn't decode its received symbol correctly, the system outage is only calculated based on the S-D link performance, which is
\begin{eqnarray}\label{eqA7}
\mathcal{P}^{\mathrm{FD}}_{\mathrm{Non-FW}}&=&\mathrm{Pr}\left[C\left(P_{\mathrm{S}}|h_{\mathrm{S,D}}|^{2}\right)< R\right]\nonumber\\
&=&\int^{e^{R}-1}_{0}f_{\mathrm{X}}(x)dx\nonumber\\
&=&1-e^{-\frac{e^{R}-1}{X}}.
\end{eqnarray}

Then, according to \eqref{eqA5}, \eqref{eqA6}, \eqref{eqA7}, and $\mathcal{P}_{\mathrm{C}}$, the system total outage probability in \textit{Proposition 1} can be obtained.

%%%%%%%%%%%%%%%%%%%%%%%%%%%%%%%%%%%%%%%%%
\bibliography{mybib}
\bibliographystyle{IEEEtran}
\end{document}